\documentclass{JHEP3}
\usepackage[pdftex]{graphicx}
\pdfoutput=1 
\usepackage{array}
\usepackage{multirow}
\usepackage{amssymb}
\newcommand{\Zll}{\mbox{$ Z\to \ell^{+} \ell^{-}$}}
\newcommand{\Zgammall}{\mbox{$Z/\gamma^{*}\to \ell^{+} \ell^{-}$}}
\newcommand{\Mll}{\mbox{$M_{\ell\ell}$}}
\newcommand{\pT}{\mbox{$p_{{}_{\scriptstyle\mathrm T}}$}}
\newcommand{\pTel}{\mbox{$p_{{}_{\scriptstyle\mathrm T\,\ell}}$}}

\newcommand{\ETel}{\mbox{$E_{{}_{\scriptstyle\mathrm T}}^{\,\ell}$}}
\newcommand{\etael}{\mbox{$\eta_{{}_{\,\scriptstyle\ell}}$}}
\newcommand{\etal}{{\it et al.}}
\newcommand{\Wml}{\mbox{$ W^{-}\to l^{-}  \overline{\nu}$}}
\newcommand{\Wpl}{\mbox{$ W^{+}\to l^{+}  \nu$}}
\date{July, 2010}
\preprint{ }
\title{Theoretical Uncertainties in Electroweak Boson Production Cross Sections at 7, 10, and 14 TeV at the LHC}
\author{N. Adam, V. Halyo, \\
    Princeton University, Princeton, NJ 08544, USA}
\author{S.A.\ Yost \\The Citadel, 171 Moultrie St., Charleston, SC 29409, USA}
\abstract{
We present an updated study of the systematic errors in the measurements 
of the electroweak boson cross-sections at the LHC for various 
experimental cuts for a center of mass energy of 7, 10 and 14 TeV.
The size of both electroweak and NNLO QCD contributions are estimated, 
together with the systematic error from the parton distributions.  The 
effects of new versions of the MSTW, CTEQ-TEA, and NNPDF PDFs are considered.
} 

\begin{document}

\section{Introduction}
A precise measurement of electroweak gauge boson production for $pp$ 
scattering will be essential in LHC physics. These processes are 
important as standard candles in the luminosity measurement, as well as 
precision electroweak parameter measurements, constraints on the PDFs, 
and as backgrounds to new physics.  The high luminosity at the LHC 
insures that systematic errors will play a dominant role in determining 
the accuracy of the cross-section. Previously,~\cite{ZStudy,WStudy} we 
studied these systematic errors at 14 TeV, focusing on missing 
electroweak corrections, NNLO QCD, and PDF uncertainties. In this update, 
we calculate the same effects for the inital LHC start-up 
energy of 7 TeV, and also an intermediate scale of 10 TeV. We also specialize 
the calculation to various experimental cuts which better reflect cuts similar 
to the one considered by the CMS and ATLAS collaborations.

We also revisit the PDF errors, because updated
PDFs have since been released which permit a significant reduction in the PDF 
errors quoted in our prevous studies.  Finally, HORACE has been updated since
the study of Ref.~\cite{WStudy}, requiring the missing electroweak contribution
to be recalculated in the $W$ case.\footnote{We thank 
P. Harris, K. Sung, P. Everaerts, and K. Hahn for bringing this to our 
attention.} This study compares the MSTW2008\cite{MSTW} (an update of
MRST\cite{MRST}), NNPDF\cite{NNPDF}, and the CTEQ-TEA family of PDF 
sets\cite{CTEQ,CT10}.~\footnote{We thank C.-P. Yuan for 
helpful discussions and for providing us with the CT10, CT10W PDF sets.}

\section{Theoretical Calculations and Monte Carlo Generators} 

The processes of interest in this paper are electroweak boson production, 
predominantly via the Drell-Yan process\cite{DY}, in which a quark and 
antiquark annihilate, producing a vector boson which decays into a lepton pair. 
The cross section mediated by a gauge boson $B$ may be inferred from the
number $N^{\rm obs}_B$ of observed events via the relation
\begin{equation}
N^{\rm obs}_B = \sigma_{\rm tot}(pp\to B\to\ell\ell') A_B\int{\cal L} dt,
\end{equation}
where $A_B$ is the acceptance obtained after applying the experimental
selection criteria. For example, if the cuts require $\pTel > \pT_{\min}$,
$0<\eta^{\ell} <\eta_{\max}$ for both leptons, then
\begin{eqnarray}
A_B(\pT_{\min},\eta_{\max}) &=& \frac{1}{\sigma_{\rm tot}(pp\to B\to\ell\ell')}
\int_{\pT_{\min}}^{\sqrt{s}/2} 
d\pTel
\int_{\pT_{\min}}^{\sqrt{s}/2} 
d\pT_{\ell'} 
\int_0^{\eta_{\max}} d\eta_{\ell}
\int_0^{\eta_{\max}} d\eta_{\ell'} 
\nonumber\\
& & \frac{d^4\sigma}{d\pTel\,d\pT_{\ell'} d\eta_{\ell}d\eta_{\ell'}}
(B\to\ell\ell')\,.
\label{eq:acc}
\end{eqnarray}
The neutral bosons $\gamma, Z$ are normally combined, and 
both $\sigma_{\rm tot}$ and the acceptance may also include
further cuts on the invariant mass $M_{\ell\ell'}$ of the lepton pair, to
prevent the cross section from being dominated by photons, which give a
divergent contribution at low energies.

The corrected $Z$  or $W$ yield
can be used as a standard candle\cite{candle} for a luminosity monitor by 
calculating the cross section and solving for $\int{\cal L} dt $.
The cross section may be calculated
by convoluting a parton-level cross section
${\widehat\sigma}_{ab}$ for partons $a,b$ with their parton
density functions (PDFs) $f_a, f_b$, giving a cross section
\begin{equation}
\sigma = \sum_{a,b} \int_0^1 
dx_1 dx_2 f_a(x_1)f_b(x_2) {\widehat\sigma}_{ab}(x_1,x_2), 
\label{eq:thxs}
\end{equation}
where the integral is over momentum fractions $x_1, x_2$.  Theoretical
uncertainties arise from limitations in the order of the calculation
of $\sigma_{ab}$, on its completeness
(for example, on whether it electroweak corrections or
$\gamma^*/Z$ interference, and on whether any phase space variables or spins
have been averaged), and on errors in the PDFs.

The simulation of hadronic processes requires an event generator 
incorporating a parton shower and hadronization. HERWIG~\cite{herwig}, 
Pythia~\cite{pythia}, and Sherpa~\cite{sherpa} are examples.  At present, 
event generators are available which incorporate NLO QCD, including 
MC@NLO~\cite{MCNLO} and POWHEG~\cite{POWHEG}. Differential NNLO 
corrections are available in vrap~\cite{vrap} FEWZ,~\cite{FEWZ} and 
DYNNLO~\cite{DYNNLO}, but 
not yet interfaced to a shower. The Resbos~\cite{resbos} family of MC programs 
calculates resummed cross-sections which improve the results in certain 
parts of phase space where large logarithms render a fixed-order calculation
unreliable. Resbos is based on a resummed NLO calculation with the addition of
approximate NNLO corrections to the Drell-Yan process. As in 
Refs.~\cite{ZStudy, WStudy}, we take MC@NLO as our basic generator, and 
calculate NNLO corrections using FEWZ, since it provides a complete calculation 
at NNLO, and the perturbative series converges well for the cuts of interest 
here. POWHEG could be used as an alternative to MC@NLO, or 
DYNNLO as an alternative to FEWZ, at the same order in QCD. 

It has been stressed that to achieve 1\% accuracy in the Drell-Yan process,
electroweak corrections must be added as well, including mixed QCD and QED
corrections, and included coherently in a single event 
generator.~\cite{WZGRAD,zykunov,QCED,Balossini,dittmaier} 
Electroweak corrections have been calculated at order $\alpha$ and
combined with a hadronic shower in HORACE.~\cite{HORACE} Another program 
incorporating electroweak corrections into a hadronic generator which has
appeared recently is WINHAC\cite{WINHAC}, for $W$ production, with a 
companion ZINHAC for $Z$ production under development. Earlier programs
calculating parton-level electroweak corrections at order $\alpha$ include
ZGRAD and WGRAD.~\cite{WZGRAD}
Partial electroweak corrections, specifically final-state photon 
radiation (FSR), can also be added to an existing shower via the program
PHOTOS.~\cite{PHOTOS} In our previous analyses~\cite{ZStudy,WStudy}, we 
used HORACE to estimate the effects of electroweak corrections on the 
systematic error in calculations done using MC@NLO and PHOTOS. We continue
this approach in the present study, using HORACE to estimate the electroweak
corrections beyond photon FSR.

\section{Analysis of Electro-Weak Corrections} 

As suggested in Refs.~\cite{ZStudy,WStudy}, our
best estimates of the cross-sections are obtained using MC@NLO with final
state radiation added via PHOTOS. 
We will begin our analysis of theoretical errors by examining the electro-weak
corrections.  The choices of cuts for this study are shown in 
Table \ref{table:cuts}. 
The results are shown in Table 
\ref{table:cs}. For the $W$, acceptances in this table are defined
relative to the total cross section, while for the $Z$, they are defined 
relative to the total cross section with an invariant mass cut 
40 GeV$ < M_Z c^2 < $ 1000 GeV to avoid
energies dominated by soft photons.

\TABLE{
\begin{tabular}{|l|lll|}
\multicolumn{4}{c}{\ }\\
\hline
Process & \multicolumn{3}{c|}{Cut}\\
\hline
$Z$ Production & 
$\ETel > 20$ GeV, & $|\etael| < 2.5$, & $70 < \Mll < 110$ GeV/$c^2$ 
\\[5pt]
$W$ Production& 
 $\ETel > 30$ GeV, & $|\etael| < 2.5$ &  \\
\hline
\end{tabular}
\caption{Acceptance regions for the Electroweak vector bosons. }
\label{table:cuts}
}

\TABLE{
\begin{tabular}{|c|c|c|c|c|}
\multicolumn{5}{c}{Cross Sections}\\
\hline
Energy & \hfil & \hfil$Z$\hfil & \hfil$W^+$\hfil & \hfil$W^-$\hfil\\
\hline
\multirow{3}{*}{7 TeV} & $\sigma_{\rm tot}$ (pb) & $1071^{+16}_{-18}\pm 2$ & $5947^{+99}_{-80} \pm 12$ & $4166^{+48}_{-98} \pm 8$ \\[5pt]
& $\sigma_{\rm cut}$ (pb) & $446.6^{+13.4}_{-8.9} \pm 1.4$ & $2440^{+22}_{-63}\pm 8$ & $1763^{+17}_{-50} \pm 5$ \\[5pt]
& $A$ & $0.417^{+0.008}_{-0.003} \pm 0$ & $0.410^{+0.001}_{-0.007} \pm 0$ & $0.423^{+0.003}_{-0.005} \pm 0$\\[5pt]
\hline
\multirow{3}{*}{10 TeV}& $\sigma_{\rm tot}$ (pb) & $1590^{+14}_{-30}\pm 3$ & $8590^{+135}_{-144} \pm 17$ & $6255^{+80}_{-148} \pm 13$ \\[5pt]
& $\sigma_{\rm cut}$ (pb) & $619.4^{+5.4}_{-22.0} \pm 2.0$ & $3219^{+48}_{-71}\pm 11$ & $2503^{+40}_{-68} \pm 8$ \\[5pt]
& $A$ & $0.389^{+0.002}_{-0.008} \pm 0$ & $0.375^{+0.003}_{-0.004} \pm 0$ & $0.400^{+0.005}_{-0.004} \pm 0$\\[5pt]
\hline
\multirow{3}{*}{14 TeV}& $\sigma_{\rm tot}$ (pb) & $2279^{+18}_{-51}\pm 5$ & $12019^{+165}_{-223} \pm 24$ & $9047^{+83}_{-247} \pm 18$ \\[5pt]
& $\sigma_{\rm cut}$ (pb) & $832.9^{+13.5}_{-21.0} \pm 2.8$ & $4179^{+46}_{-115}\pm 14$ & $3455^{+39}_{-115} \pm 11$ \\[5pt]
& $A$ & $0.366^{+0.007}_{-0.003} \pm 0$ & $0.348^{+0.003}_{-0.007} \pm 0$ & $0.382^{+0.003}_{-0.005} \pm 0$\\[5pt]
\hline
\end{tabular}
\caption{Total and cut cross sections $\sigma_{\rm tot}, \sigma_{\rm cut}$ 
and acceptances $A$ for the cuts in 
Table \ref{table:cuts} using MC@NLO and PHOTOS. The errors are separated into 
PDF uncertainties (MSTW2008), with upper and lower bounds, and MC uncertainties.}
\label{table:cs}
}

As in Refs.~\cite{ZStudy, WStudy}, we used HORACE~\cite{HORACE} to 
study the error arising from missing ${\cal O}(\alpha)$ electroweak (EWK)
corrections.  This program includes initial and final-state QED radiation in 
a photon shower approximation and exact ${\cal O}(\alpha)$ EWK
corrections matched to a leading-log QED shower. Specifically, the 
missing EWK contribution was calculated by generating events for $pp 
\to Z/\gamma^* \to \ell^{+}\ell^{-}$ and $pp \to W^{\pm} \to 
\ell^{\pm}\nu$ using HORACE with its full {\cal O}($\alpha$) corrections
and parton-showered with HERWIG. These were compared to events generated
again by HORACE, but without EWK corrections (Born-level), also showered
with HERWIG, and HERWIG plus PHOTOS. MSTW2008 PDFs were used in
the calculations. 

\TABLE{
\begin{tabular}{|l|l|c|c|c|c|}
\multicolumn{5}{c}{\bf Electro-Weak Corrections}\\
\multicolumn{5}{c}{ }\\
\multicolumn{5}{c}{  $Z$ Production}\\
\hline
Energy &  & Born & Born+FSR  & Electro-Weak & Difference \\
\hline
\multirow{3}{*}{7 TeV} 
& $\sigma_{\rm tot}$
& $906.47 \pm 0.40$    & $906.47\pm 0.40$    
& $922.14 \pm 1.04$    & $+1.70 \pm 0.12$\% \\ 
& $\sigma_{\rm cut}$
& $356.72 \pm 0.46$    & $333.60  \pm 0.48$    
& $332.82 \pm 0.50$    & $-0.23 \pm 0.21$\% \\
& $A$
& $0.3935 \pm 0.0005$  & $0.3680 \pm 0.0006$ 
& $0.3609 \pm 0.0007$  & $+1.96 \pm 0.24$\% \\
\hline
\multirow{3}{*}{10 TeV} 
& $\sigma_{\rm tot}$
& $1359.25 \pm 0.80$    & $1359.25 \pm 0.80$    
& $1387.67 \pm 1.09$    & $+2.05 \pm 0.10$\% \\ 
& $\sigma_{\rm cut}$
& $494.58 \pm 0.63$    & $462.65  \pm 0.66$    
& $462.38  \pm 0.68$   & $-0.06 \pm 0.20$\% \\
& $A$
& $0.3639 \pm 0.0005$   & $0.3404\pm 0.0005$ 
& $0.3332 \pm 0.0006$  & $+2.15 \pm 0.23$\% \\
\hline
\multirow{3}{*}{14 TeV} 
& $\sigma_{\rm tot}$
& $1964.76 \pm 1.13$    & $1964.76 \pm 1.13$    
& $2001.20 \pm 1.79$    & $+1.82 \pm 0.10$\% \\ 
& $\sigma_{\rm cut}$
& $669.09 \pm 0.86$    & $625.66  \pm 0.89$    
& $625.97  \pm 0.89$   & $+0.05 \pm 0.20$\% \\
& $A$
& $0.3405 \pm 0.0005$  & $0.3184 \pm 0.0005$ 
& $0.3128 \pm 0.0005$  & $+1.81 \pm 0.23$\% \\
\hline
\multicolumn{5}{c}{~}\\
\multicolumn{5}{c}{  $W^+$ Production}\\
\hline
Energy& & Born & Born+FSR & Electro-Weak & Difference\\
\hline
\multirow{3}{*}{7 TeV} 
& $\sigma_{\rm tot}$
&$ 4993.2 \pm 0.4 $&$ 4993.2 \pm 0.4 $
&$ 4948.5 \pm 0.3 $&$ -0.904 \pm 0.009 $\% \\ 
& $\sigma_{\rm cut}$
&$ 2065 \pm 5  $&$  1940 \pm 5 $
&$ 1932 \pm 5  $&$ -0.41 \pm 0.36 $\% \\
& $A$
&$ 0.4136 \pm 0.0010 $&$ 0.3885 \pm 0.0010 $
&$ 0.3904 \pm 0.0010 $&$  +0.49 \pm 0.36 $\% \\
\hline
\multirow{3}{*}{10 TeV} 
& $\sigma_{\rm tot}$
&$ 7271.1  \pm 0.7 $&$  7271.1 \pm 0.7 $
&$ 7205.6 \pm 0.5 $&$ -0.899 \pm 0.014 $\% \\ 
& $\sigma_{\rm cut}$
&$ 2748 \pm 7 $&$  2579 \pm 7 $
&$ 2566 \pm 7 $&$ -0.60 \pm 0.45 $\% \\
& $A$
&$ 0.3780 \pm 0.0010 $&$ 0.3547 \pm 0.0010 $
&$ 0.3561 \pm 0.0010 $&$  +0.30 \pm 0.37 $\% \\
\hline
\multirow{3}{*}{14 TeV} 
& $\sigma_{\rm tot}$
&$ 10384 \pm 1 $&$ 10384 \pm 1 $
&$ 10350 \pm 1 $&$ -0.322 \pm 0.014 $\% \\ 
& $\sigma_{\rm cut}$
&$ 3575 \pm 10 $&$  3372 \pm 10 $&
$ 3350 \pm 10  $&$ -0.68 \pm 0.41 $\% \\
& $A$
&$ 0.3443 \pm 0.0010 $&$ 0.3248 \pm 0.0009 $
&$ 0.3236 \pm 0.0009 $&$  -0.36 \pm 0.41 $\% \\
\hline
\multicolumn{5}{c}{~}\\
\multicolumn{5}{c}{  $W^-$ Production}\\
\hline
Energy &  & Born & Born+FSR & Electro-Weak & Difference \\
\hline
\multirow{3}{*}{7 TeV} 
& $\sigma_{\rm tot}$
&$ 3535.2 \pm 0.2 $&$ 3535.2 \pm 0.2 $
&$ 3504.0 \pm 0.2 $&$-0.890 \pm 0.008 $\% \\ 
& $\sigma_{\rm cut}$
&$ 1489 \pm 4 $&$ 1412 \pm 3 $
&$ 1397 \pm 3 $&$-1.03  \pm 0.35 $\% \\
& $A$
&$ 0.4213 \pm 0.0010 $&$ 0.3993 \pm 0.0010 $
&$ 0.3987 \pm 0.0010 $&$-0.14 \pm 0.35 $\% \\
\hline
\multirow{3}{*}{10 TeV} 
& $\sigma_{\rm tot}$
&$ 5355.6 \pm 0.7 $&$ 5355.6 \pm 0.7 $
&$ 5307.9 \pm 0.3 $&$-0.908  \pm 0.012 $\% \\ 
& $\sigma_{\rm cut}$
&$ 2114 \pm 5 $&$ 2001 \pm 5 $
&$ 1989 \pm 5 $&$-0.51  \pm 0.39 $\% \\
& $A$
&$ 0.3948 \pm 0.0010 $&$ 0.3735 \pm 0.0010 $
&$ 0.3747 \pm 0.0010 $&$+0.39 \pm 0.39 $\% \\
\hline
\multirow{3}{*}{14 TeV} 
& $\sigma_{\rm tot}$
&$ 7899.2 \pm 0.8  $&$ 7899.2 \pm 0.8 $
&$ 7875.7 \pm 0.6  $&$-0.297  \pm 0.013 $\% \\ 
& $\sigma_{\rm cut}$
&$ 2919 \pm 8 $&$ 2747 \pm 8 $
&$ 2748 \pm 8 $&$+0.03  \pm 0.39 $\% \\
& $A$
&$ 0.3695 \pm 0.0010 $&$ 0.3477 \pm 0.0010 $
&$ 0.3489 \pm 0.0010 $&$+0.32 \pm 0.39 $\% \\
\hline
\end{tabular}
\caption{Electro-Weak corrections to the $pp\to Z \to \ell^{+}\ell^{-}$, 
$pp\to W^+ \to \ell^{+}\nu_{\ell}$, and $pp \to W^- \to \ell^{-}\bar\nu_{\ell}$ 
total and cut cross-sections $\sigma_{\rm tot}, \sigma_{\rm cut}$ and 
acceptance $A$. The Born + FSR column shows 
corrections obtained using PHOTOS, and the Electro-Weak column shows corrections
generated using HORACE 3.1, for $\ell=e$ or $\mu$.  The difference is the 
missing electroweak correction if PHOTOS is used instead of HORACE, with the 
correct sign. }
\label{table:horace_xs_comp}
}

The results are shown in Table \ref{table:horace_xs_comp}. The acceptance is 
the ratio of the cut to total cross section. In the $Z$ case, each
total cross section entry includes an invariant mass cut 40 GeV $< M_{Z}c^2 <
1000$ GeV, as noted above.  Final state 
radiation should dominate for both $W$ and $Z$ 
production,\cite{WZGRAD} so there is good justification for passing the 
Born-level events through PHOTOS to add final-state photons, as is
verified in previous LHC studies.\cite{LESHOUCHES, ZStudy}. 

Figs.~\ref{fig:wp_fsr}, \ref{fig:wm_fsr} and the the $W^{\pm}$ acceptance 
results in Table \ref{table:horace_xs_comp} show the improvement to the
kinematic distributions obtained by using PHOTOS.  Table 
\ref{table:horace_xs_comp} shows that the error introduced by using 
PHOTOS in place of the more complete electroweak corrections of HORACE is 
typically less than 1\%, both in the cross sections and acceptances. 
Larger errors found in Ref.\ \cite{WStudy} were due to a technical
problem in the earlier version of HORACE, which has been corrected in 
the current version 3.1 used in the present study.

\section{QCD Errors: NNLO Corrections and Scale Dependence} 

The QCD error is divided into two categories, the error due to stopping at 
NLO order, and the dependence on residual dependence on the factorization
and renormalization scales, due to truncating the calculation at finite order.
For an estimate of the missing higher-order corrections, we take the NNLO 
calculation obtained using the current state-of-the-art program, 
FEWZ~\cite{FEWZ}. The residual scale dependence is estimated by calculating
the NNLO result and varying the factorization and renormalization scales. 
The NNLO scale dependence is relevant here, 
since we may consider the NNLO result to be known, but neglected. 

To obtain the results shown, we ran FEWZ~\cite{FEWZ} using using 
MSTW2008\cite{MSTW} PDFs at the appropriate QCD order, and calculated
the difference between the NLO and NNLO result, choosing the renormalization 
and factorization scales at $M_Z$ or $M_W$, as appropriate. The estimates of
residual scale dependence at NNLO was found by repeating the calculations
with the renormalization and factorization scales multiplied by 2 or 1/2. Both
scales are taken to be equal.

The size of the NNLO contribution, as well as the residual scale
dependence at order NNLO, is shown in table \ref{table:qcd}. The scale
dependence is calculated by dividing the standard deviation of the
results at the three scales by their average.  If the NLO result at the 
center scale is taken as the result, an estimate of the combined error is
shown in the final column, adding the two errors in quadrature as 
in Refs.~\cite{ZStudy,WStudy}. 

\TABLE{
\begin{tabular}{|l|l|c|c|}
\multicolumn{4}{c}{\bf NNLO QCD Corrections and Scale Dependence}\\
\multicolumn{4}{c}{ }\\
\multicolumn{4}{c}{  $Z$ Production}\\
\hline
Energy &  & NNLO QCD (\%) & QCD Scale (\%)  \\
\hline
\multirow{3}{*}{7 TeV} 
& $\sigma_{\rm tot}$   & $+2.86 \pm 0.10$ & $0.42 \pm 0.02$\% \\ 
& $\sigma_{\rm cut}$   & $+2.74 \pm 0.62$ & $0.21 \pm 0.08$\% \\
& $A$                  & $-0.12 \pm 0.61$ & $0.54 \pm 0.21$\% \\
\hline
\multirow{3}{*}{10 TeV} 
& $\sigma_{\rm tot}$   & $+2.62 \pm 0.10$ & $0.53\pm 0.03$ \\
& $\sigma_{\rm cut}$   & $+2.01 \pm 0.73$ & $0.59\pm 0.26$ \\
& $A$                  & $-0.60 \pm 0.71$ & $1.05\pm 0.46$ \\
\hline
\multirow{3}{*}{14 TeV} 
& $\sigma_{\rm tot}$   & $+2.24 \pm 0.11$ & $0.57 \pm 0.03$\% \\ 
& $\sigma_{\rm cut}$   & $+2.40  \pm 0.71$ & $1.10 \pm 0.44$\% \\
& $A$                  & $+0.16 \pm 0.70$  & $0.54 \pm 0.22$\% \\
\hline
\multicolumn{4}{c}{~}\\
\multicolumn{4}{c}{  $W^+$ Production}\\
\hline
Energy &  & NNLO QCD (\%) & QCD Scale (\%)  \\
\hline
\multirow{3}{*}{7 TeV} 
& $\sigma_{\rm tot}$   & $ +3.09 \pm 0.25 $&$ 0.41 \pm 0.08 $\% \\ 
& $\sigma_{\rm cut}$   & $ +3.35 \pm 0.70 $&$ 0.47 \pm 0.19 $\% \\
& $A$                  & $ +0.25 \pm 0.72 $&$ 0.75 \pm 0.34 $\% \\
\hline
\multirow{3}{*}{10 TeV} 
& $\sigma_{\rm tot}$   &$ +2.80 \pm 0.35 $&$ 0.70 \pm 0.13 $\% \\ 
& $\sigma_{\rm cut}$   &$ +3.47 \pm 0.64 $&$ 0.35 \pm 0.14 $\% \\
& $A$                  &$ +0.65 \pm 0.71 $&$ 0.40 \pm 0.18 $\% \\
\hline
\multirow{3}{*}{14 TeV} 
& $\sigma_{\rm tot}$   &$ +2.11 \pm 0.39 $&$ 0.047\pm 0.020$\% \\ 
& $\sigma_{\rm cut}$   &$ +3.86 \pm 0.77 $&$ 1.41 \pm 0.58$\% \\
& $A$                  &$ +1.71 \pm 0.85 $&$ 1.45 \pm 0.88$\% \\
\hline
\multicolumn{4}{c}{~}\\
\multicolumn{4}{c}{  $W^-$ Production}\\
\hline
Energy &  & NNLO QCD (\%) & QCD Scale (\%)  \\
\hline
\multirow{3}{*}{7 TeV} 
& $\sigma_{\rm tot}$  &$ +2.69 \pm 0.18 $&$ 0.54 \pm 0.09$\% \\ 
& $\sigma_{\rm cut}$  &$ +2.80 \pm 0.66 $&$ 0.59 \pm 0.23$\% \\
& $A$                 &$ +0.11 \pm 0.67 $&$ 0.84 \pm 0.35$\% \\
\hline
\multirow{3}{*}{10 TeV} 
& $\sigma_{\rm tot}$  &$ +2.31 \pm 0.51 $&$ 0.88 \pm 0.23$\% \\ 
& $\sigma_{\rm cut}$  &$ +3.83 \pm 0.62 $&$ 0.24 \pm 0.08$\% \\
& $A$                 &$ +1.48 \pm 0.79 $&$ 1.08 \pm 0.45$\% \\
\hline
\multirow{3}{*}{14 TeV} 
& $\sigma_{\rm tot}$  &$ +2.20 \pm 0.38 $&$ 0.87 \pm 0.18$\% \\ 
& $\sigma_{\rm cut}$  &$ +3.52 \pm 0.63 $&$ 0.86 \pm 0.30$\% \\
& $A$                 &$ +1.30 \pm 0.72 $&$ 0.093\pm 0.038$\% \\
\hline
\end{tabular}
\caption{QCD errors for the total cross and cut cross sections 
$\sigma_{\rm tot}, \sigma_{\rm cut}$ and acceptance $A$ using the cuts from 
Table \ref{table:cuts}. The NNLO contribution obtained using 
FEWZ\cite{FEWZ} includes the correct sign, and could be added as a $K$-factor 
if desired. The scale dependence is calculated at three scales, $M_B$, $2M_B$, 
and $M_B/2$, for the appropriate boson mass $M_B$.  The standard deviation of 
the three results is divided by their average to obtain the scale dependence 
shown. }
\label{table:qcd}
}

\section{Parton Distribution Function Errors} 

Since our previous studies,\cite{ZStudy,WStudy}, 
new PDF sets have been released incorporating the latest data.  We consider the
most recent PDF sets from three groups: MSTW~\cite{MSTW}, 
NNPDF\cite{NNPDF}, and CTEQ-TEA~\cite{CT10}.  PDF errors for MSTW and 
CT10 were calculated as in Refs.~\cite{ZStudy,WStudy} using the
asymmetric Hessian method, where the cross-section results from the 
various eigenvector PDF sets have been combined according to the 
prescriptions found in Refs.~\cite{CTEQ,MRST,MSTW}. NNPDF is based on 
an entirely different neural-network based approach, which provides a 
collection of replica PDFs, for which the standard deviation provides a
symmetric error estimate.

Tables \ref{table:pdf-Z} -- \ref{table:pdf-W-} compare the 
cross sections, acceptances, and their attendant uncertainties for four sets 
of PDFs: MSTW2008\cite{MSTW} NNPDF2.0\cite{NNPDF}, and both CT10 and 
CT10W\cite{CT10} from the latest CTEQ-TEA analysis.  The two CT10 sets differ 
in their treatment of the Tevatron D0 Run-2 $W$ lepton asymmetry 
data.\cite{Wasym} Since this data creates tension 
with existing constraints on $d(x)/u(x)$, two approaches were taken: CT10 omits 
this data, while CT10W includes it with an extra weight.\cite{CT10} All of the 
PDF sets in this comparison are NLO versions with 90\% CL.  The cuts are as in 
Table \ref{table:cuts}, and the upper and lower limits of the asymmetric errors 
are shown. (Symmetric errors are given for NNPDF.) The results for 7 TeV are 
also shown in Figs.~\ref{fig:pdf_xs_Z} --~\ref{fig:pdf_xs_Wminus}.

In addition we studied the sensitivity of the kinematic acceptance calculations
to the uncertainties affecting the PDF sets. The errors considered here are due
to the experimental inputs used to obtain the PDFs, not to the QCD order of the
calculation, which should be matched to the order of the hard matrix element 
used.  Figs.~\ref{fig:pdf_xs_vs_cut_Z} --~\ref{fig:pdf_xs_vs_cut_Wminus}
show the systematic error on the production
cross-sections as a function of the $|\eta|$ cut, minimum
lepton $\pT$ cut and (where appropriate) $M_{T}$ cut for the values given in 
Table\ \ref{table:cuts}.  The fractional uncertainties, shown in
in the same figures, demonstrate that
the relative uncertainty in the cross-section is very flat as a function
of the kinematic cuts, until the region of extreme cuts and low
statistics in the MC are reached. The corresponding uncertainty on the
acceptance as a function of the kinematic cuts is shown in
Figs~\ref{fig:pdf_acc_vs_cut_Z} --~\ref{fig:pdf_acc_vs_cut_Wminus}. 
These show a similar dependence to the cross-section uncertainties,
though the fractional errors are smaller.

\TABLE{
\begin{tabular}{|c|rrr|rrr|rrr|}
\multicolumn{10}{c}{\bf PDF Errors in $Z$ Production}\\
\multicolumn{10}{c}{ }\\
\multicolumn{10}{c}{7 TeV}\\
\hline
PDF Set &
$\sigma_{\rm tot}$&$\Delta\sigma_{\rm tot}^+$&$\Delta\sigma_{\rm tot}^-$&
$\sigma_{\rm cut}$&$\Delta\sigma_{\rm cut}^+$&$\Delta\sigma_{\rm cut}^-$&
$A$ & $\Delta A^+$ & $\Delta A^-$\\
\hline
MSTW2008
&1071 $\pm$ 2 &16 &18 	&447 $\pm$ 1 &13 &9 	&0.417 &0.008 &0.003 \\
NNPDF
&1032 $\pm$ 2 &22 &22 	&429 $\pm$ 1 &7 &7	&0.415 &0.005 &0.005 \\
CT10
&1068 $\pm$ 2 &32 &35 	&447 $\pm$ 1 &20 &15 	&0.418 &0.009 &0.003 \\
CT10W
&1072 $\pm$ 2 &27 &39 	&449 $\pm$ 1 &13 &20 	&0.418 &0.006 &0.006 \\
\hline
\multicolumn{10}{c}{ }\\
\multicolumn{10}{c}{10 TeV}\\
\hline
PDF Set & 
$\sigma_{\rm tot}$&$\Delta\sigma_{\rm tot}^+$&$\Delta\sigma_{\rm tot}^-$&
$\sigma_{\rm cut}$&$\Delta\sigma_{\rm cut}^+$&$\Delta\sigma_{\rm cut}^-$&
$A$ & $\Delta A^+$ & $\Delta A^-$\\
\hline
MSTW2008
&1590 $\pm$ 3 &15 &30 	&619 $\pm$ 2 &5 &22 	&0.389 &0.002 &0.008 \\
NNPDF
&1524 $\pm$ 3 &30 &30 	&619 $\pm$ 2 &10 &10 	&0.387 &0.005 &0.005 \\
CT10
&1585 $\pm$ 3 &61 &44 	&620 $\pm$ 2 &27 &27 	&0.391 &0.006 &0.009 \\
CT10W
&1589 $\pm$ 3 &55 &51 	&618 $\pm$ 2 &30 &20 	&0.389 &0.009 &0.004 \\
\hline
\multicolumn{10}{c}{ }\\
\multicolumn{10}{c}{14 TeV}\\
\hline
PDF Set & 
$\sigma_{\rm tot}$&$\Delta\sigma_{\rm tot}^+$&$\Delta\sigma_{\rm tot}^-$&
$\sigma_{\rm cut}$&$\Delta\sigma_{\rm cut}^+$&$\Delta\sigma_{\rm cut}^-$&
$A$ & $\Delta A^+$ & $\Delta A^-$\\
\hline
MSTW2008
&2279 $\pm$ 5 &18 &51 	&833 $\pm$ 3 &14 &21 	&0.366 &0.007 &0.003 \\
NNPDF
&2174 $\pm$ 4 &40 &40 	&796 $\pm$ 3 &16 &16 	&0.366 &0.005 &0.005 \\
CT10
&2286 $\pm$ 5 &68 &97 	&844 $\pm$ 3 &25 &50 	&0.369 &0.004 &0.009 \\
CT10W
&2287 $\pm$ 5 &73 &96 	&836 $\pm$ 3 &43 &29 	&0.366 &0.012 &0.002 \\
\hline
\end{tabular}
\caption{PDF Errors in the cross sections ($\sigma$) and acceptances ($A$) 
for the cuts in Table \ref{table:cuts}, calculated using the asymmetric 
Hessian method for MSTW and CT10(W), and the standard deviation for the NNPDF 
replica sets.  All cross sections are in pb, while the acceptance is a ratio 
of cut to total cross sections. Statistical MC errors are shown for the 
$\sigma_{tot}, \sigma_{cut}$ entries, and are 
zero for the acceptances $A$, to the precision shown.} 
\label{table:pdf-Z}
}

\TABLE{
\begin{tabular}{|c|rrr|rrr|rrr|}
\multicolumn{10}{c}{\bf PDF Errors in $W^+$ Production}\\
\multicolumn{10}{c}{ }\\
\multicolumn{10}{c}{7 TeV}\\
\hline
PDF Set &
$\sigma_{\rm tot}$&$\Delta\sigma_{\rm tot}^+$&$\Delta\sigma_{\rm tot}^-$&
$\sigma_{\rm cut}$&$\Delta\sigma_{\rm cut}^+$&$\Delta\sigma_{\rm cut}^-$&
$A$ & $\Delta A^+$ & $\Delta A^-$\\
\hline
MSTW2008
&5947 $\pm$ 12 &99 &80 	 &2440 $\pm$ 8 &22 &63 	&0.410 &0.001 &0.007 \\
NNPDF
&5794 $\pm$ 12 &146 &146 &2376 $\pm$ 7 &53 &53 	&0.410 &0.002 &0.002 \\
CT10
&5993 $\pm$ 12 &178 &223 &2464 $\pm$ 8 &58 &121 &0.411 &0.002 &0.008 \\
CT10W
&5968 $\pm$ 12 &209 &190 &2454 $\pm$ 8 &100 &67 &0.411 &0.004 &0.003 \\
\hline
\multicolumn{10}{c}{ }\\
\multicolumn{10}{c}{10 TeV}\\
\hline
PDF Set & 
$\sigma_{\rm tot}$&$\Delta\sigma_{\rm tot}^+$&$\Delta\sigma_{\rm tot}^-$&
$\sigma_{\rm cut}$&$\Delta\sigma_{\rm cut}^+$&$\Delta\sigma_{\rm cut}^-$&
$A$ & $\Delta A^+$ & $\Delta A^-$\\
\hline
MSTW2008
&8590 $\pm$ 17 &135 &144 &3219 $\pm$ 11 &48 &71  &0.375 &0.003 &0.004 \\
NNPDF
&8311 $\pm$ 17 &200 &200 &3124 $\pm$ 10 &65 &65  &0.376 &0.003 &0.003 \\
CT10
&8644 $\pm$ 17 &348 &267 &3246 $\pm$ 11 &137 &98 &0.376 &0.004 &0.003 \\
CT10W
&8629 $\pm$ 17 &304 &293 &3253 $\pm$ 11 &149 &126 &0.377 &0.007 &0.006 \\
\hline
\multicolumn{10}{c}{ }\\
\multicolumn{10}{c}{14 TeV}\\
\hline
PDF Set & 
$\sigma_{\rm tot}$&$\Delta\sigma_{\rm tot}^+$&$\Delta\sigma_{\rm tot}^-$&
$\sigma_{\rm cut}$&$\Delta\sigma_{\rm cut}^+$&$\Delta\sigma_{\rm cut}^-$&
$A$ & $\Delta A^+$ & $\Delta A^-$\\
\hline
MSTW2008
&12019 $\pm$ 24 &165 &223 &4179 $\pm$ 14 &46 &115  &0.348 &0.003 &0.007 \\
NNPDF
&11559 $\pm$ 23 &267 &267 &4032 $\pm$ 14 &82 &82   &0.349 &0.003 &0.003 \\
CT10
&12159 $\pm$ 24 &366 &493 &4231 $\pm$ 14 &159 &159 &0.348 &0.006 &0.003 \\
CT10W
&12095 $\pm$ 24 &453 &420 &4222 $\pm$ 14 &190 &135 &0.349 &0.006 &0.003 \\
\hline
\end{tabular}
\caption{PDF Errors in the cross sections ($\sigma$) and acceptances ($A$) 
for the cuts in Table \ref{table:cuts}, calculated using the asymmetric 
Hessian method for MSTW and CT10(W), and the standard deviation for the NNPDF 
replica sets.  All cross sections are in pb, while the acceptance is a ratio 
of cut to total cross sections. Statistical MC errors are shown for the 
$\sigma_{tot}, \sigma_{cut}$ entries, and are 
zero for the acceptances $A$, to the precision shown.} 
\label{table:pdf-W+}
}

\TABLE{
\begin{tabular}{|c|rrr|rrr|rrr|}
\multicolumn{10}{c}{\bf PDF Errors in $W^-$ Production}\\
\multicolumn{10}{c}{ }\\
\multicolumn{10}{c}{7 TeV}\\
\hline
PDF Set &
$\sigma_{\rm tot}$&$\Delta\sigma_{\rm tot}^+$&$\Delta\sigma_{\rm tot}^-$&
$\sigma_{\rm cut}$&$\Delta\sigma_{\rm cut}^+$&$\Delta\sigma_{\rm cut}^-$&
$A$ & $\Delta A^+$ & $\Delta A^-$\\
\hline
MSTW2008
&4166 $\pm$ 8 &48 &98 	&1763 $\pm$ 5 &17 &50 	&0.423 &0.003 &0.005 \\
NNPDF
&3943 $\pm$ 8 &89 &89 	&1648 $\pm$ 5 &30 &30 	&0.418 &0.004 &0.004 \\
CT10
&4084 $\pm$ 8 &141 &147	&1723 $\pm$ 5 &51 &76 	&0.422 &0.002 &0.008 \\
CT10W
&4117 $\pm$ 8 &165 &121	&1702 $\pm$ 5 &114 &14 	&0.413 &0.017 &0.001 \\
\hline
\multicolumn{10}{c}{ }\\
\multicolumn{10}{c}{10 TeV}\\
\hline
PDF Set & 
$\sigma_{\rm tot}$&$\Delta\sigma_{\rm tot}^+$&$\Delta\sigma_{\rm tot}^-$&
$\sigma_{\rm cut}$&$\Delta\sigma_{\rm cut}^+$&$\Delta\sigma_{\rm cut}^-$&
$A$ & $\Delta A^+$ & $\Delta A^-$\\
\hline
MSTW2008
&6255 $\pm$ 13 &80 &148  &2503 $\pm$ 8 &40 &68 	&0.400 &0.005 &0.004 \\
NNPDF
&5900 $\pm$ 12 &123 &123 &2338 $\pm$ 7 &40 &40 	&0.396 &0.004 &0.004 \\
CT10
&6164 $\pm$ 12 &211 &248 &2460 $\pm$ 8 &81 &105 &0.399 &0.005 &0.006 \\
CT10W
&6214 $\pm$ 12 &219 &241 &2456 $\pm$ 8 &97 &92 &0.395 &0.004 &0.003 \\
\hline
\multicolumn{10}{c}{ }\\
\multicolumn{10}{c}{14 TeV}\\
\hline
PDF Set & 
$\sigma_{\rm tot}$&$\Delta\sigma_{\rm tot}^+$&$\Delta\sigma_{\rm tot}^-$&
$\sigma_{\rm cut}$&$\Delta\sigma_{\rm cut}^+$&$\Delta\sigma_{\rm cut}^-$&
$A$ & $\Delta A^+$ & $\Delta A^-$\\
\hline
MSTW2008
&9047 $\pm$ 18 &83 &247  &3456 $\pm$ 11 &39 &115 &0.382 &0.003 &0.005 \\
NNPDF
&8490 $\pm$ 17 &166 &166 &3218 $\pm$ 10 &55 &55  &0.379 &0.004 &0.004 \\
CT10
&8925 $\pm$ 18 &362 &325 &3379 $\pm$ 11 &201 &75 &0.379 &0.013 &0.001 \\
CT10W
&8979 $\pm$ 18 &369 &337 &3394 $\pm$ 11 &139 &132 &0.378 &0.004 &0.003 \\
\hline
\end{tabular}
\caption{PDF Errors in the cross sections ($\sigma$) and acceptances ($A$) 
for the cuts in Table \ref{table:cuts}, calculated using the asymmetric 
Hessian method for MSTW and CT10(W), and the standard deviation for the NNPDF 
replica sets.  All cross sections are in pb, while the acceptance is a ratio 
of cut to total cross sections. Statistical MC errors are shown for the 
$\sigma_{tot}, \sigma_{cut}$ entries, and are 
zero for the acceptances $A$, to the precision shown.} 
\label{table:pdf-W-}
}

The new PDF sets have improved the error estimates relative to our 
previous studies.\cite{ZStudy,WStudy}.
Uncertainties in the cross-sections contributed by the PDFs are 
typically in the range of 1 -- 2\% for MSTW2008 and NNPDF2.0, and 2 -- 3\% for 
CT10(W). The acceptances have smaller errors in all cases, typically close to
1\%, with less dependence on the choice of PDF.  The error bands of CT10(W) 
typically encompass those of MSTW2008 and 
NNPDF2.0, and appear to be a more conservative error estimate.

\section{Summary} 

Tables \ref{table:cs-errors} and \ref{table:acc-errors} show the combined 
errors in the cross sections and acceptances, respectively. These tables 
separate the errors into three classes, and 
calculate the total errors under the assumption that they are independent. 
Since the missing NNLO and missing electroweak corrections are known, 
including the sign, these can be combined to give a total error 
for missing higher-order effects, labeled ``Higher Order'' in the 
table.\footnote{In previous studies \cite{ZStudy, WStudy}, the missing 
EWK and QCD errors were treated as independent random errors, but the 
present treatment is more accurate, since these are systematic errors of 
known sign.} The QCD scale uncertainty remaining at NNLO is included as a
separate random error. The NNLO scale dependence is relevant since the 
calculation has been done at this order, although we may choose not to 
implement the resulting $K$-factor. Symmetric errors are used for the PDFs. 
The CT10 errors are shown in these tables, since these are more conservative
than for the other PDF sets, and typically encompass the PDF differences. 
The cross section error is much more sensitive to the choice of PDFs than the 
acceptance error.  For acceptances, we have verified that CT10 and MSTW2008 
give very similar acceptance errors, in spite of significant differences 
in the cross section errors. 

\TABLE{
\begin{tabular}{|c|l|l|l|}
\multicolumn{4}{c}{\bf Summary of Cross Section Errors}\\
\multicolumn{4}{c}{\ }\\
\multicolumn{4}{c}{$Z$ Production}\\
\hline
Energy & 7 TeV & 10 TeV & 14 TeV \\
\hline
Higher Order & $+2.50\pm 0.65$  & $+1.95\pm 0.75$ & $+2.45\pm 0.73$ \\
QCD Scale & $0.21\pm 0.08$   & $0.59\pm 0.26$   & $1.10\pm 0.44$ \\
PDF & $3.69\pm 0.00$   & $4.14\pm 0.00$   & $4.11\pm 0.00$ \\
\hline 
Total & $4.46\pm 0.48$  & $4.62\pm 0.32$   & $4.91\pm 0.38$ \\ 
\hline
\multicolumn{4}{c}{\ }\\
\multicolumn{4}{c}{$W^+$ Production}\\
\hline
Energy & 7 TeV & 10 TeV & 14 TeV \\
\hline
Higher Order & $+2.94\pm 0.78$  & $+2.95\pm 0.75$  & $+3.19\pm 0.87$ \\
QCD Scale & $0.47\pm 0.19$   & $0.35\pm 0.14$   & $1.41\pm 0.58$ \\
PDF & $3.33\pm 0.00$   & $3.71\pm 0.00$   & $3.65\pm 0.00$ \\
\hline 
Total & $ 4.47\pm 0.52$   & $ 4.75\pm 0.47$   & $5.05\pm 0.58$ \\ 
\hline
\multicolumn{4}{c}{\ }\\
\multicolumn{4}{c}{$W^-$ Production}\\
\hline
Energy & 7 TeV & 10 TeV & 14 TeV \\
\hline
Higher Order & $+1.78\pm 0.75$    & $+3.23\pm 0.72$   & $+3.55\pm 0.74$ \\
QCD Scale & $0.59\pm 0.23$   & $0.24\pm 0.08$   & $0.86\pm 0.30$ \\
PDF & $3.49\pm 0.00$   & $3.48\pm 0.00$   & $3.76\pm 0.00$ \\
\hline 
Total & $3.96\pm 0.34$   & $4.76\pm 0.49$   & $5.24\pm 0.$ \\ 
\hline
\end{tabular}
\caption{Summary of systematic errors in the cross sections ($\sigma_{\rm cut}$)
for $Z$ and $W^\pm$ production for the cuts of Table \ref{table:cuts} 
at each of the three energy scales cross sections. The sign 
of the higher-order correction error shows the sign of the correction needed.}
\label{table:cs-errors}
}

\TABLE{
\begin{tabular}{|c|l|l|l|}
\multicolumn{4}{c}{\bf Summary of Acceptance Errors}\\
\multicolumn{4}{c}{\ }\\
\multicolumn{4}{c}{$Z$ Production}\\
\hline
Energy & 7 TeV & 10 TeV & 14 TeV \\
\hline
Higher Order & $-2.04\pm 0.65 $ & $-2.70\pm 0.75$ & $-1.61\pm 0.73$\\
QCD Scale & $0.54\pm 0.21$ & $1.05\pm 0.46$ & $0.54\pm 0.22$\\
PDF & $1.21\pm 0.00$ & $1.67\pm 0.00$ & $1.43\pm 0.00$\\
\hline 
Total & $2.43\pm 0.55$ & $3.35\pm 0.62$ & $2.22\pm 0.53$\\ 
\hline
\multicolumn{4}{c}{\ }\\
\multicolumn{4}{c}{$W^+$ Production}\\
\hline
Energy & 7 TeV & 10 TeV & 14 TeV \\
\hline
Higher Order & $+0.75\pm 0.81$ & $+1.04\pm 0.81$ & $+1.36\pm 0.94$\\
QCD Scale & $0.75\pm 0.34$ & $0.40\pm 0.18$ & $1.45\pm 0.88$\\
PDF & $0.89\pm 0.00$ & $0.85\pm 0.00$ & $1.21\pm 0.00$\\
\hline 
Total & $1.38\pm 0.47$ & $1.40\pm 0.60$ & $2.33\pm 0.78$\\ 
\hline
\multicolumn{4}{c}{\ }\\
\multicolumn{4}{c}{$W^-$ Production}\\
\hline
Energy & 7 TeV & 10 TeV & 14 TeV \\
\hline
Higher Order & $-0.03\pm 0.76$  & $+1.78\pm 0.87$ & $+1.62\pm 0.82$\\
QCD Scale & $0.84\pm 0.35$ & $1.08\pm 0.45$ & $0.09\pm 0.04$\\
PDF & $0.95\pm 0.00$ & $1.21\pm 0.00$ & $1.29\pm 0.00$\\
\hline 
Total & $1.27\pm 0.24$ & $2.41\pm 0.68$ & $2.07\pm 0.64$\\ 
\hline
\end{tabular}
\caption{Summary of systematic errors in the acceptance ($A$) for 
$Z$ and $W^\pm$ production for the cuts of Table \ref{table:cuts} at 
each of the three energy scales.  The sign of the higher-order correction 
error shows the sign of the correction needed.}
\label{table:acc-errors}
}

The total uncertainties are typically in the range of $4 - 5$\% for the cut 
cross sections, and $1 - 3$\% for the acceptances. The QCD errors are 
comparable to those seen in the earlier studies at 14 TeV.\cite{ZStudy, WStudy} 
There is a significant improvement in the PDF errors (see Tables~\ref{table:pdf-Z},\ref{table:pdf-W+},\ref{table:pdf-W-}), especially for acceptances, though
they still dominate the errors in the cross sections.
The higher-order electroweak contribution has been reduced 
significantly in the $W$ case, due to a correction to the HORACE program. 
Although the electroweak and NNLO QCD corrections are not yet available in an 
event generator, they can be calculated using HORACE and FEWZ, respectively, 
and used to improve the calculation via $K$-factors, reducing the quoted errors
to the sum (in quadrature) of the PDF and NNLO QCD scale uncertainties.
\footnote{However, in the case of NNLO QCD, some care is needed because the 
shower already incorporates some NNLO effects, which would partially duplicate 
those calculated by FEWZ or DYNNLO.} An implementation of the next level of 
electroweak and QCD corrections in an event generator would have the 
potential for a significant improvement.

\section*{Acknowledgments}
This work was supported in part by D.o.E. grants DE-FG02-91ER40671 and 
DE-SC0003925.  We acknowledge valuable communications with S. Frixione, 
C.-P.\ Yuan, P.M.\ Nadolsky, F. Petriello, K. Melnikov, P. Harris, K. Sung, 
P. Everaerts, and K. Hahn.  We thank C.-P.\ Yuan for making the CT10 and 
CT10W PDF sets available for this study before their official release, and 
for reviewing and discussing our PDF analysis.  S. Yost thanks 
Princeton University and CERN for support and hospitality
during portions of this work, and The Citadel Foundation for support.

\newpage

\setlength{\unitlength}{1in}
\FIGURE{
\begin{picture}(6.5,2.5)(0,0)
\put(0,0){
  \includegraphics[scale=0.4]{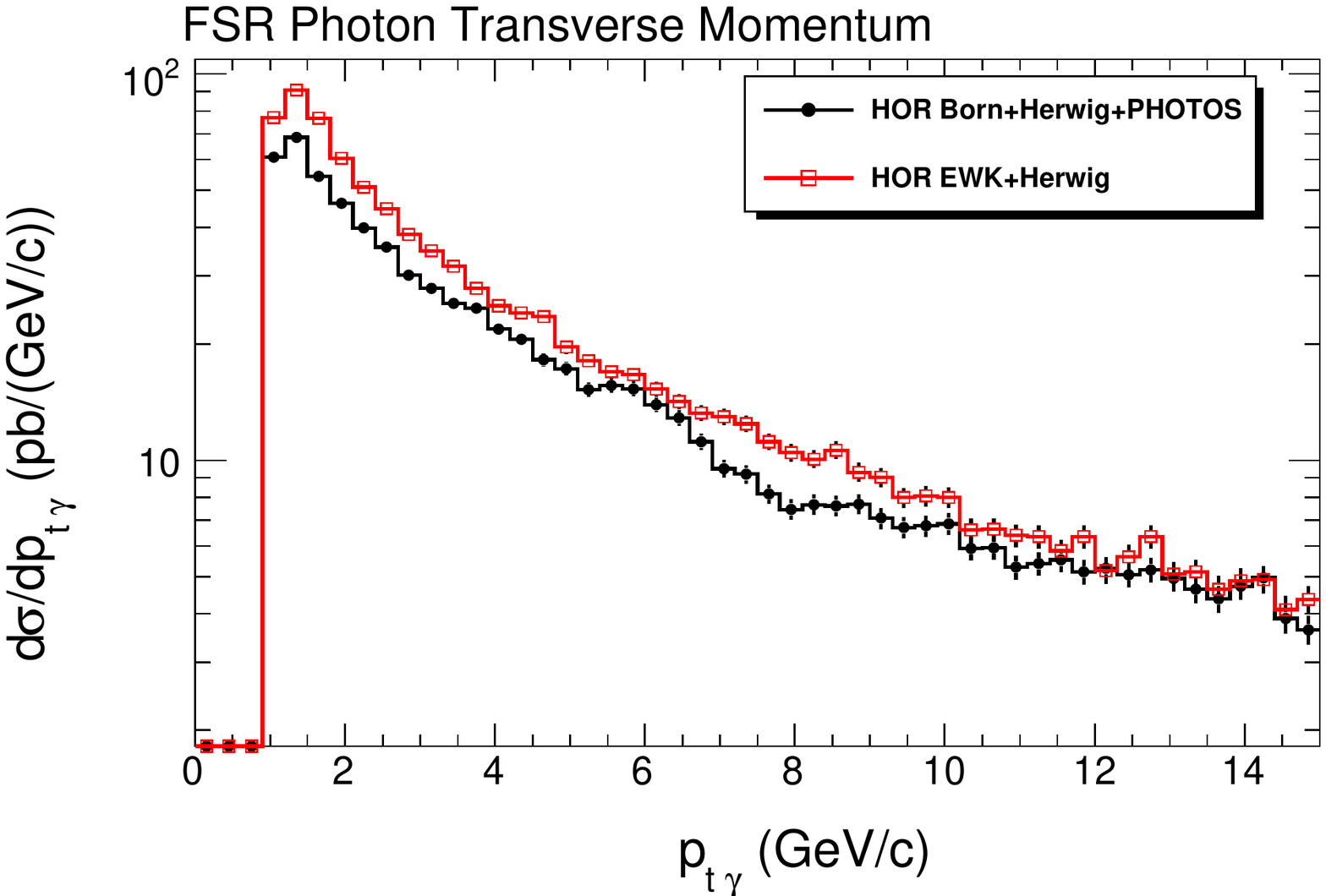}
}
\put(3,0){
  \includegraphics[scale=0.4]{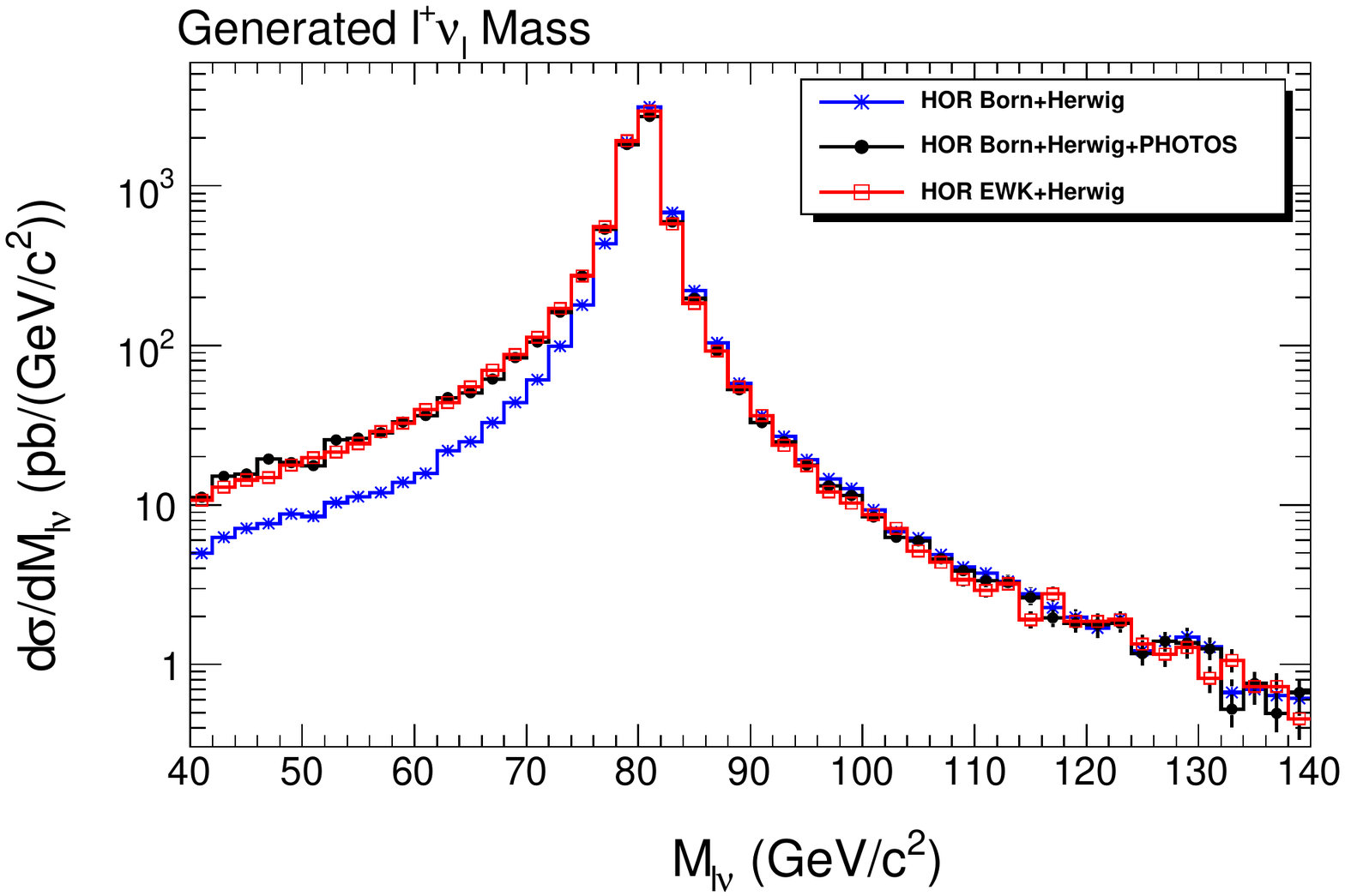}
}
\end{picture}
\caption{$W^+$: FSR photon $p_\mathrm{T}$ and electron--neutrino invariant
  mass distributions. The addition of PHOTOS to the HERWIG showering improves
  the shape of the kinematic distributions.}
\vspace{10pt}
\label{fig:wp_fsr}
}

\vfill
\FIGURE{
\begin{picture}(6.5,2.0)(0,0)
\put(0,0){ 
  \includegraphics[scale=0.4]{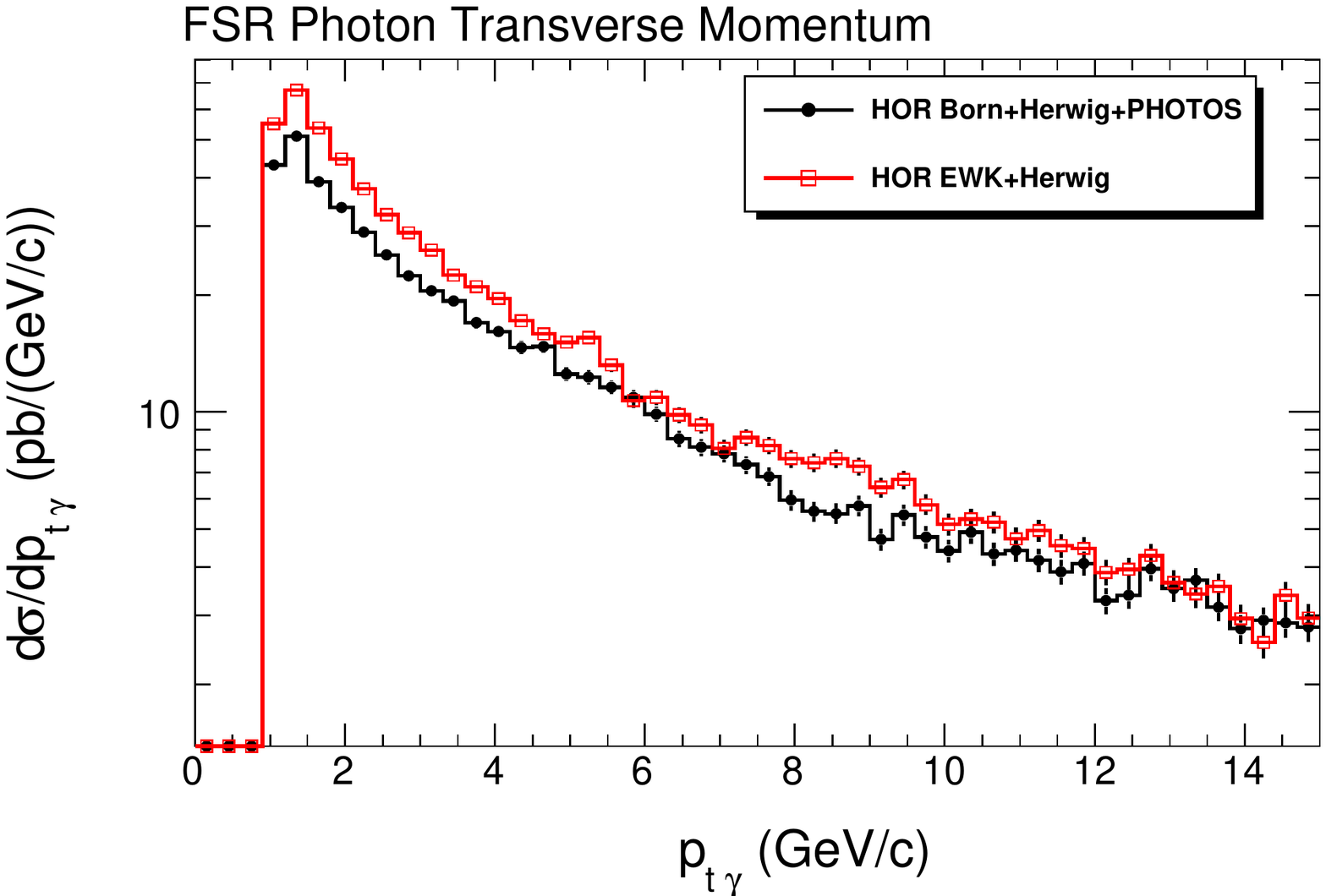}
}
\put(3,0){
  \includegraphics[scale=0.4]{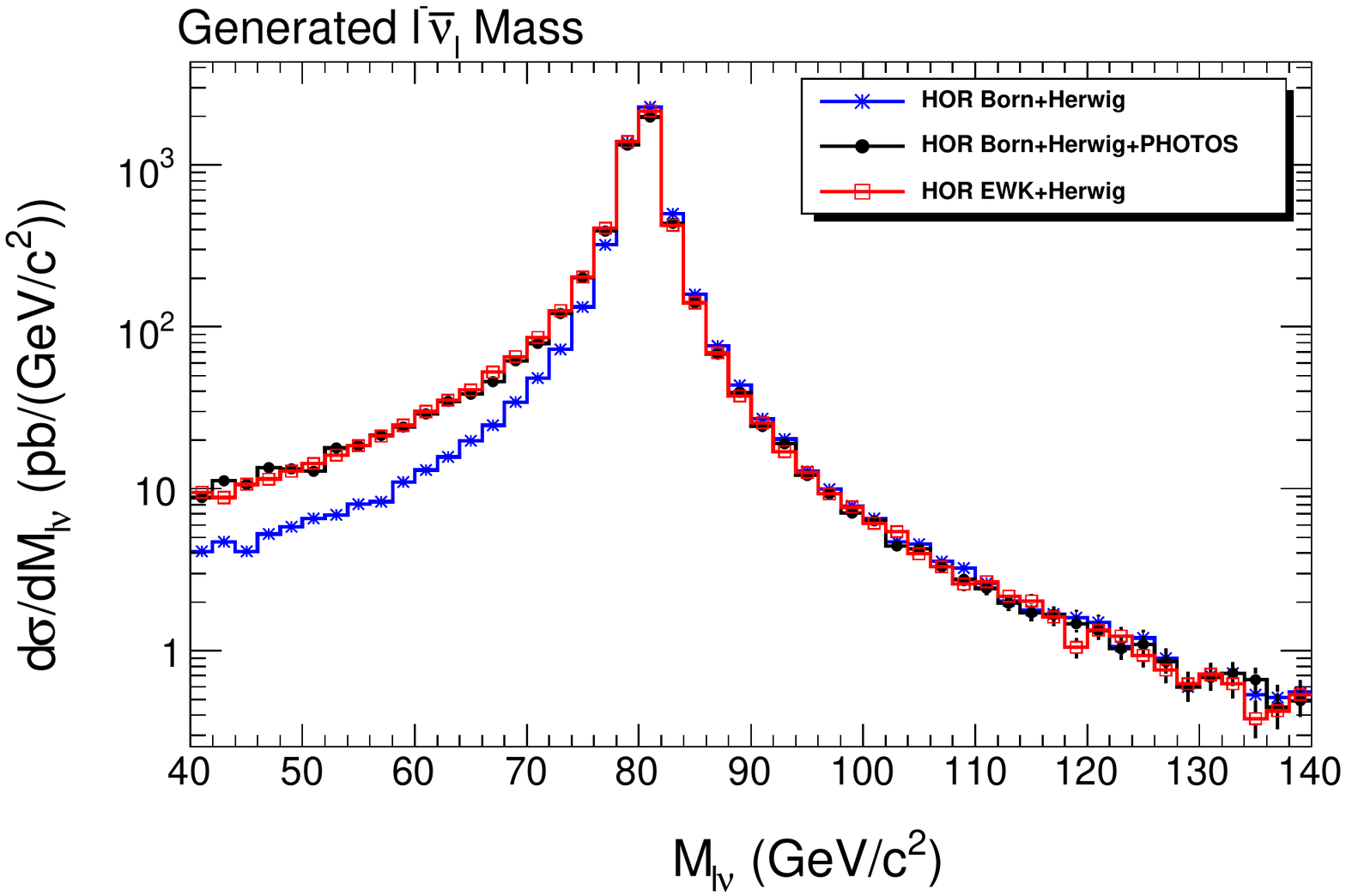}
}
\end{picture}
\caption{$W^-$: FSR photon $p_\mathrm{T}$ and electron--neutrino invariant
  mass distributions. The addition of PHOTOS to the HERWIG showering improves
  the shape of the kinematic distributions. }
\label{fig:wm_fsr}
}

\FIGURE[ht]{
\includegraphics[width=9cm]{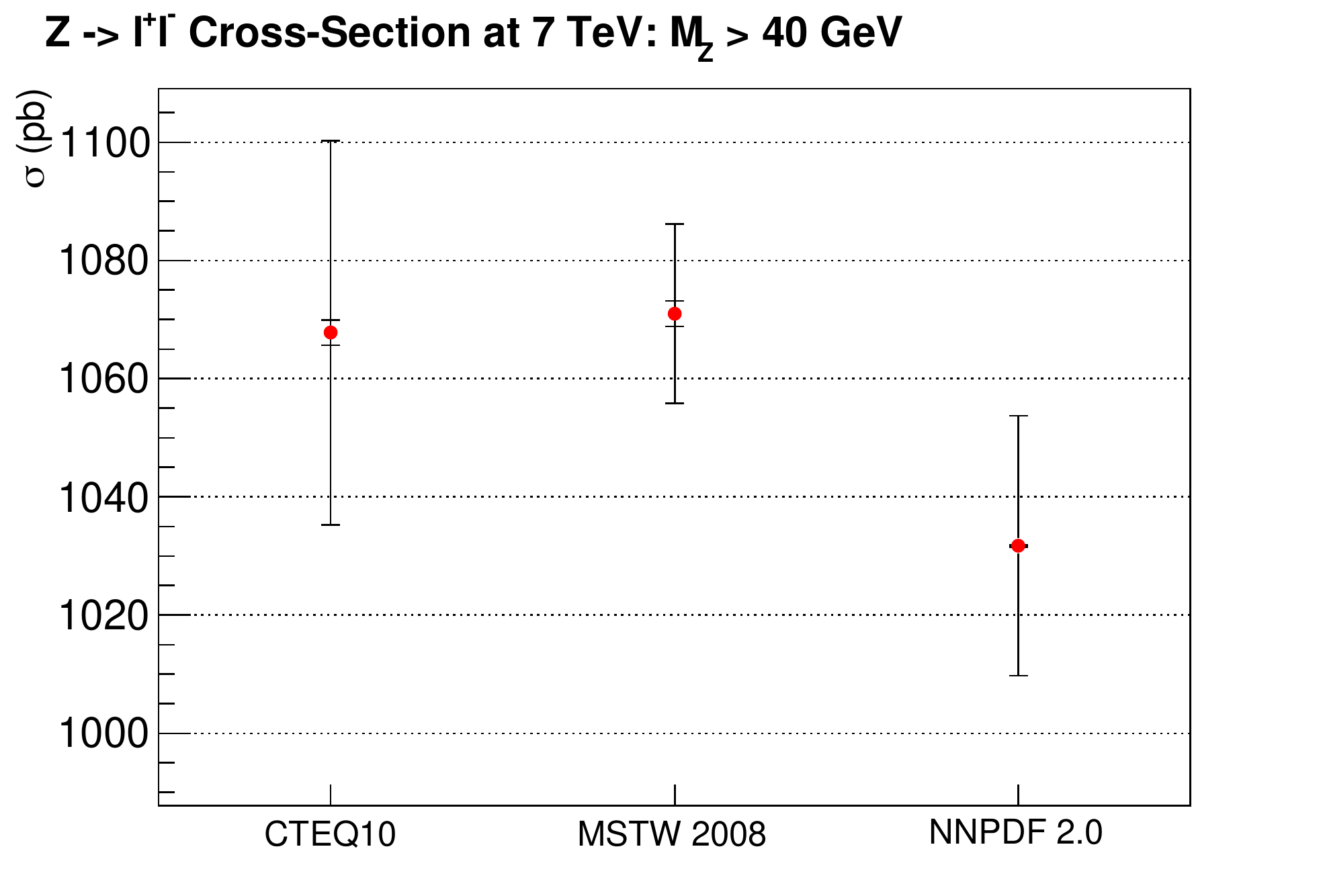}
\caption{ Comparison of $\Zgammall$ cross-sections
  for $\Mll > 40$ GeV/$c^{2}$ for several recent PDF calculations. }
\label{fig:pdf_xs_Z}
}

\FIGURE[ht]{
\includegraphics[width=9cm]{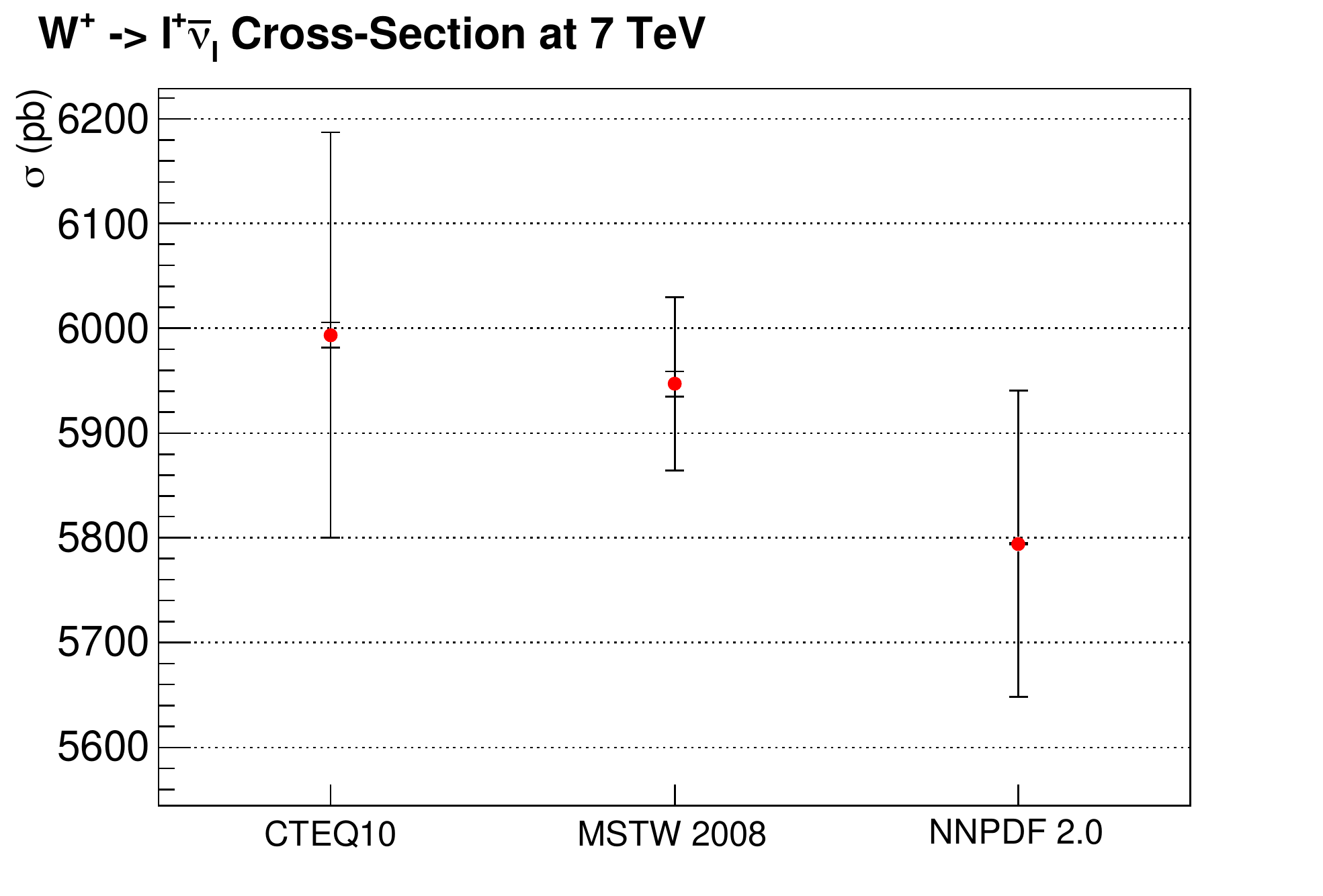}
\caption{ Comparison of $\Wpl$ total cross-sections
   for several recent PDF calculations. }
\label{fig:pdf_xs_Wplus}
}

\FIGURE[ht]{
\includegraphics[width=9cm]{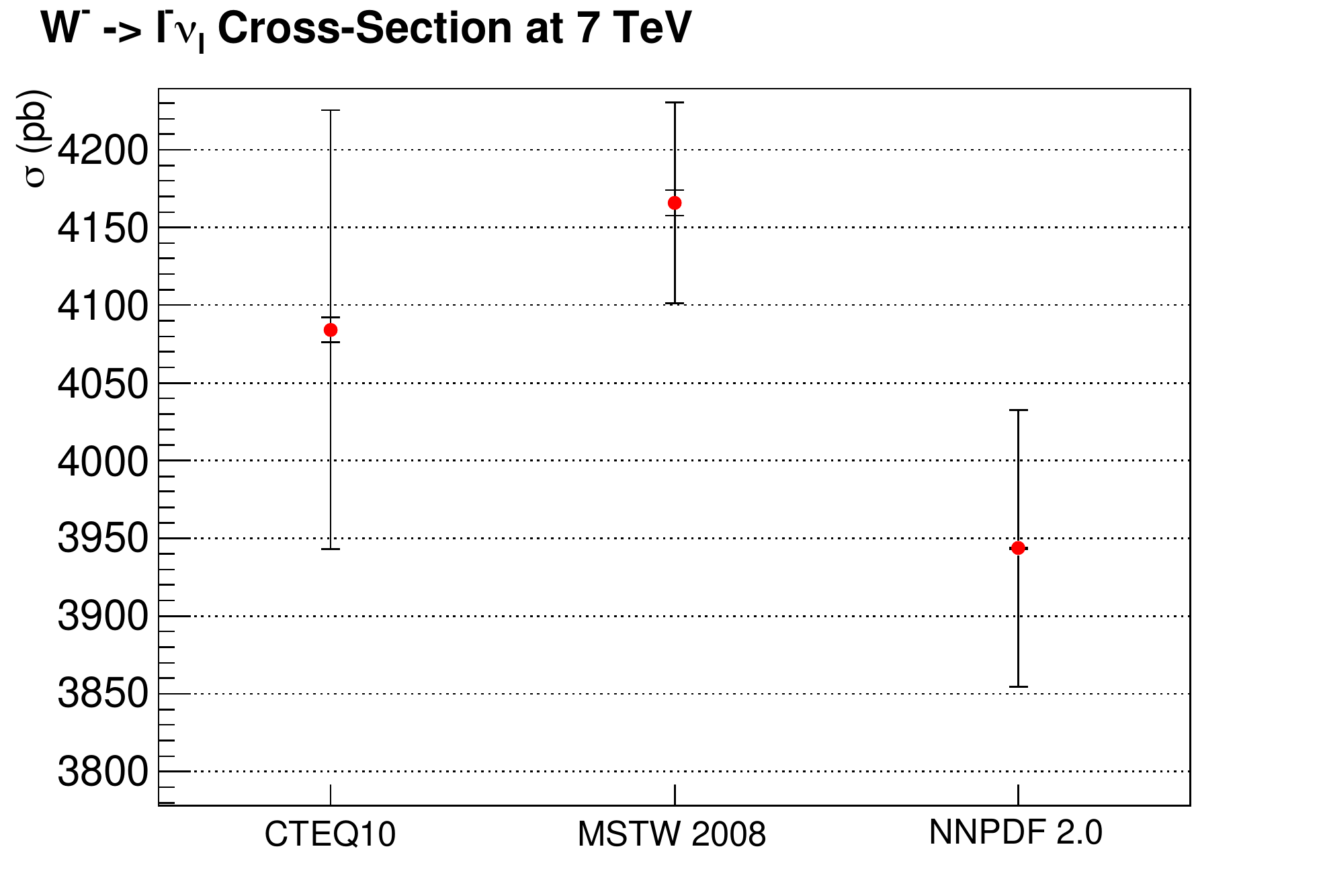}
\caption{ Comparison of $\Wml$ total cross-sections
  for several recent PDF calculations. }
\label{fig:pdf_xs_Wminus}
}

\FIGURE[ht]{
\begin{tabular}{cc}
\includegraphics[width=7cm]{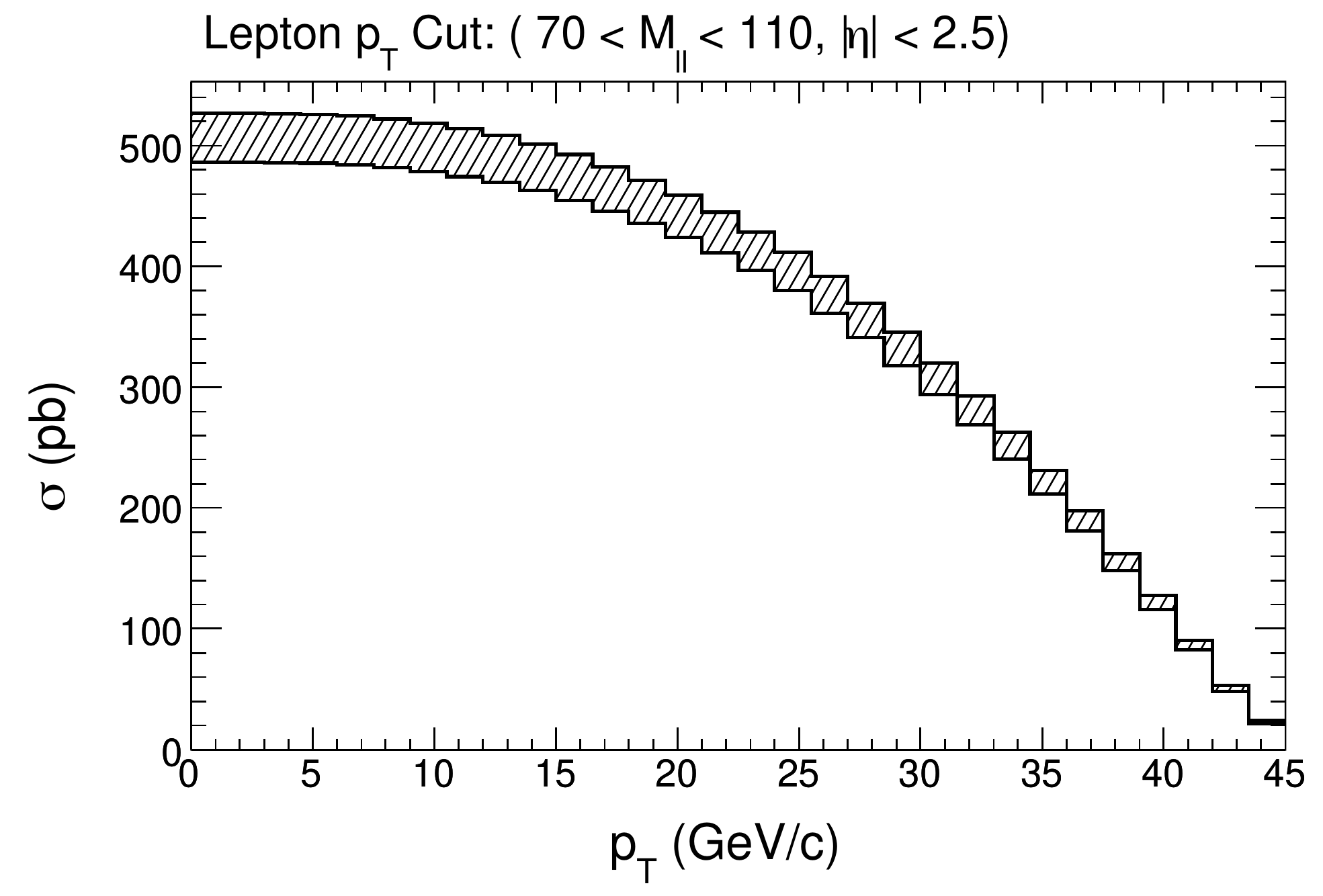} &
\includegraphics[width=7cm]{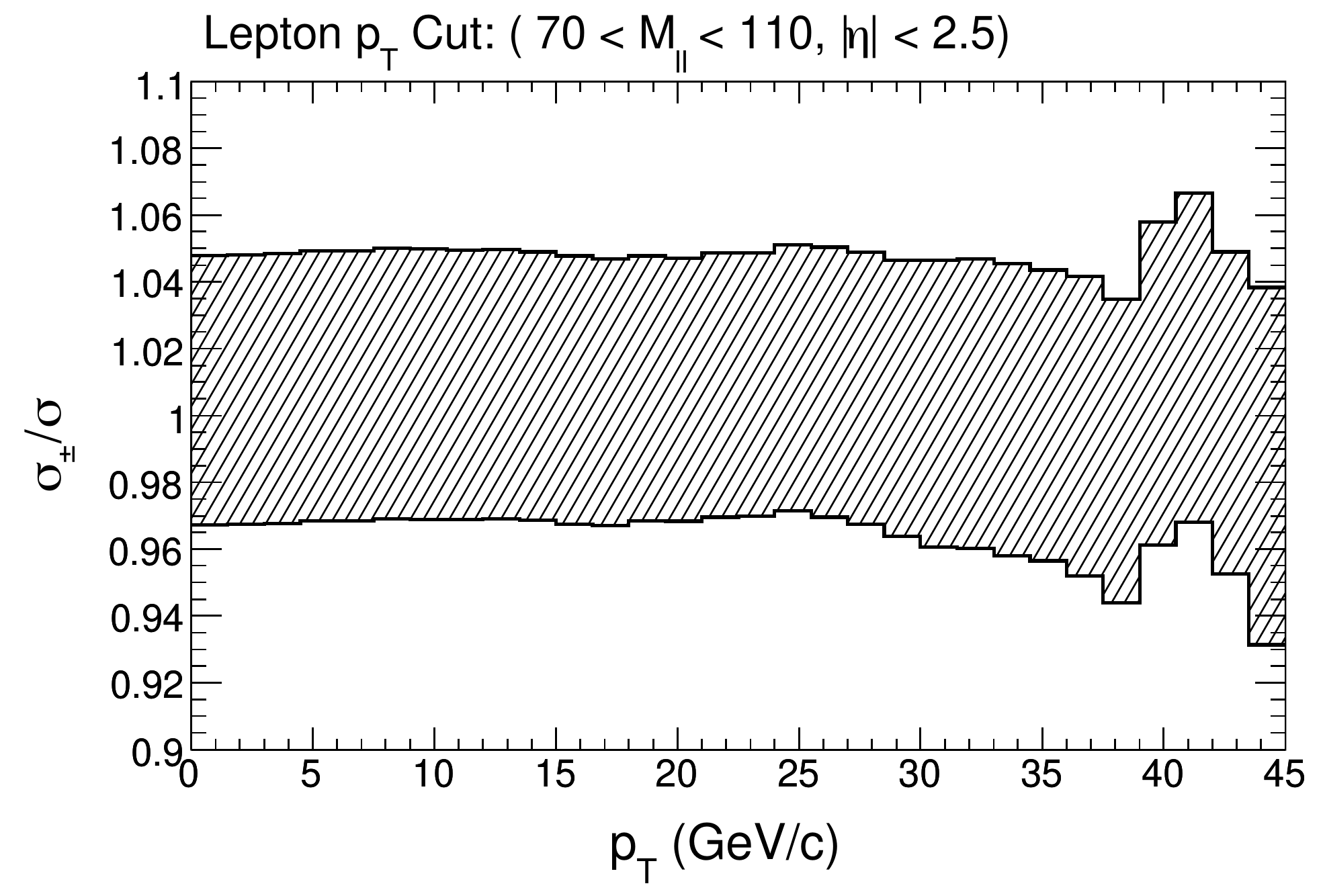} \\
\multicolumn{2}{c}{(a)} \\
\includegraphics[width=7cm]{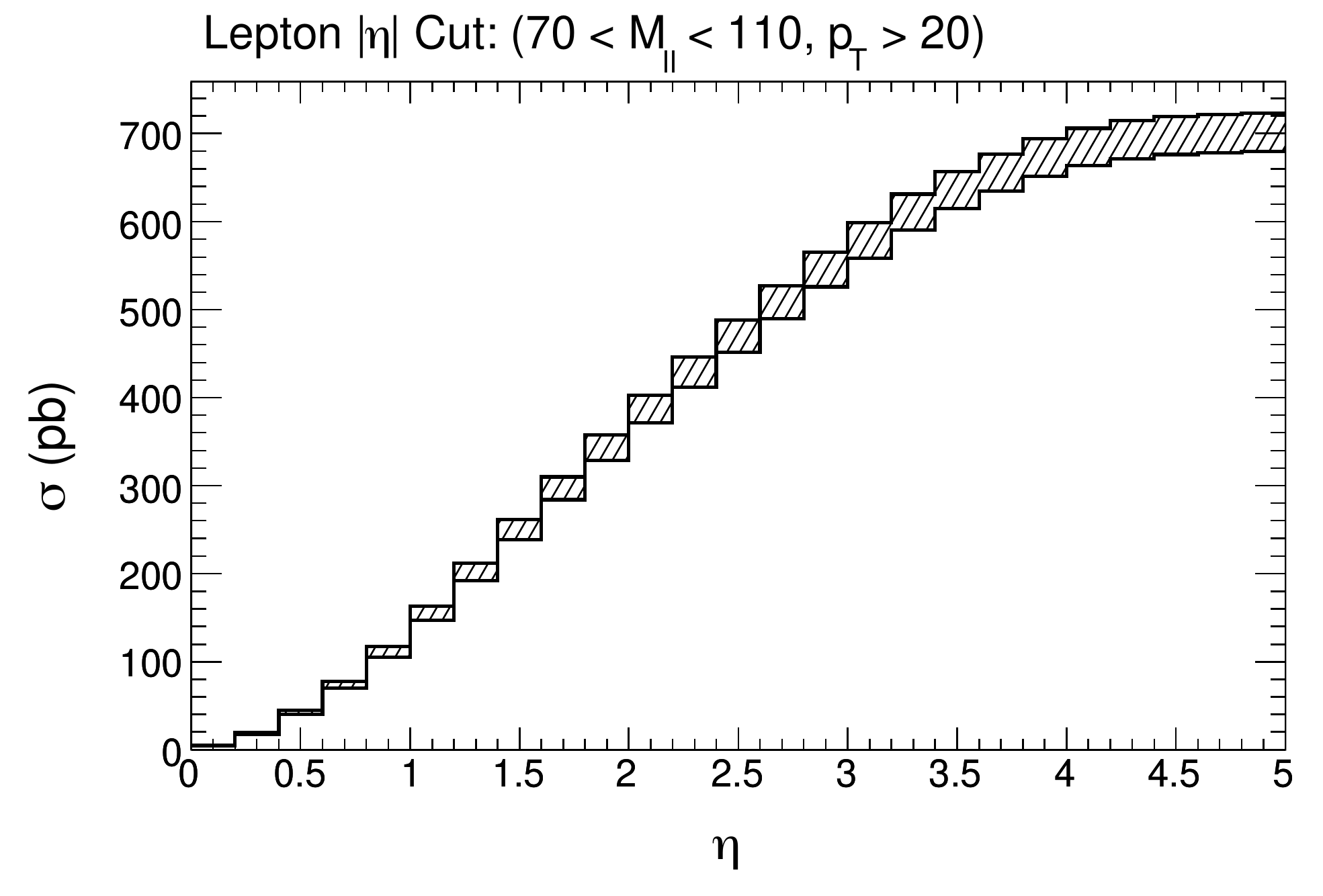}&
\includegraphics[width=7cm]{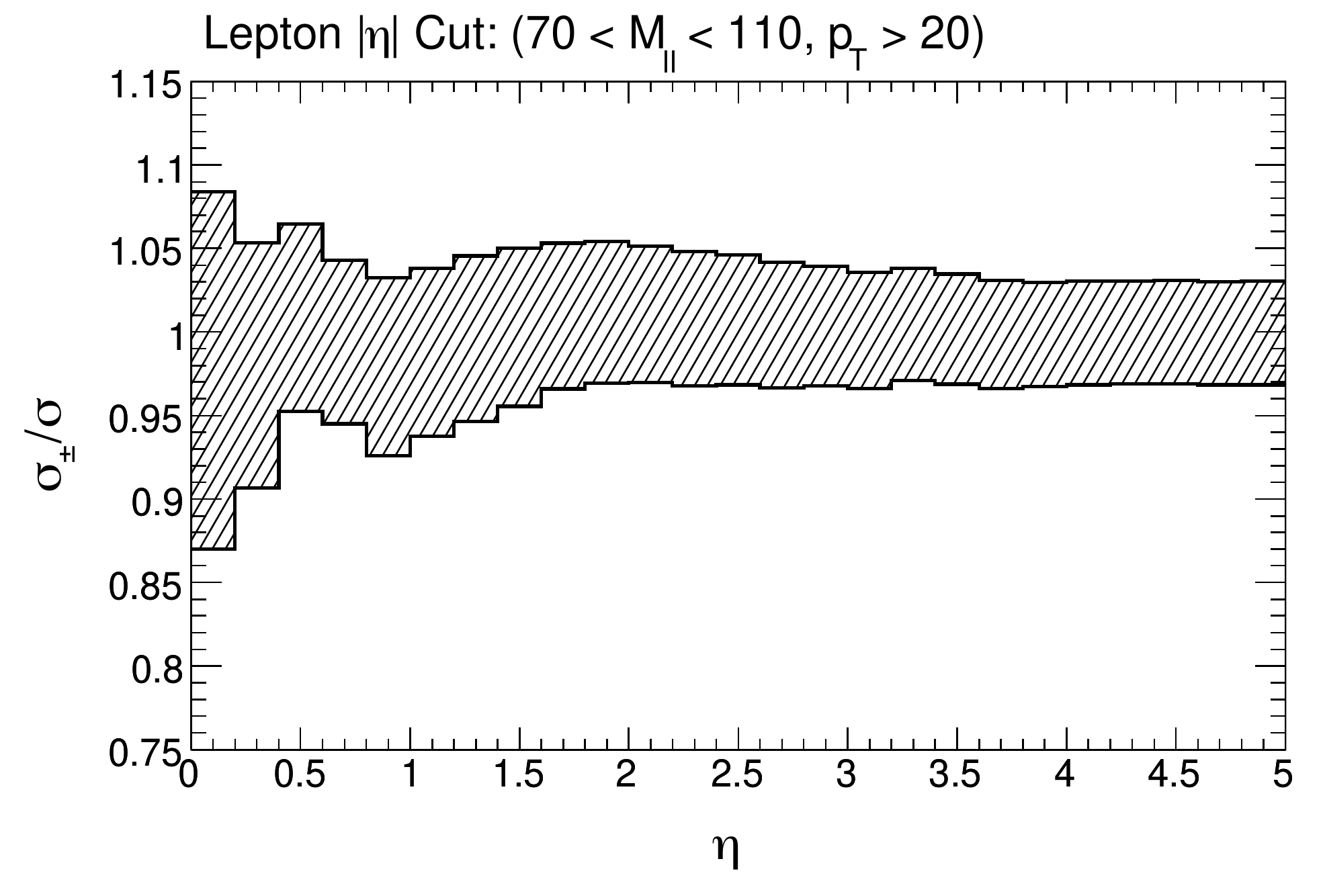}\\
\multicolumn{2}{c}{(b)} \\
\end{tabular}
\caption{ The $Z/\gamma^* \to \ell^{+}\ell^{-}$ cross-section $\sigma$ at
  7 TeV, as a
  function of the (a) $\pT$ cut and (b) $\eta$ cut for acceptance regions
  as defined in Table~\ref{table:cuts}. We fix all cuts except the
  cut to be varied at their specified values. The figures on the right
  show the relative errors in the cross sections.}
\label{fig:pdf_xs_vs_cut_Z}
}

\FIGURE[ht]{
\begin{tabular}{cc}
\includegraphics[width=7cm]{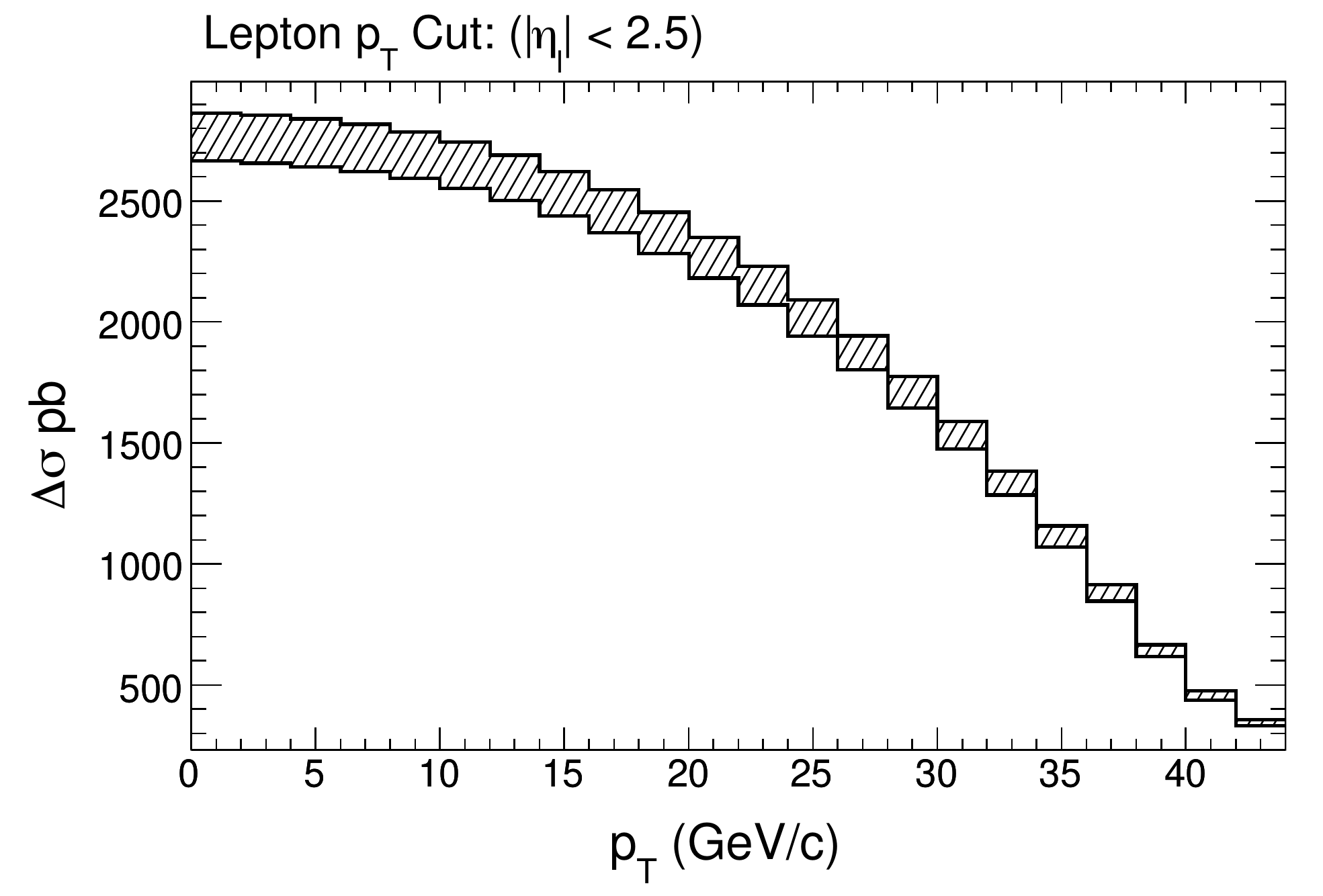} &
\includegraphics[width=7cm]{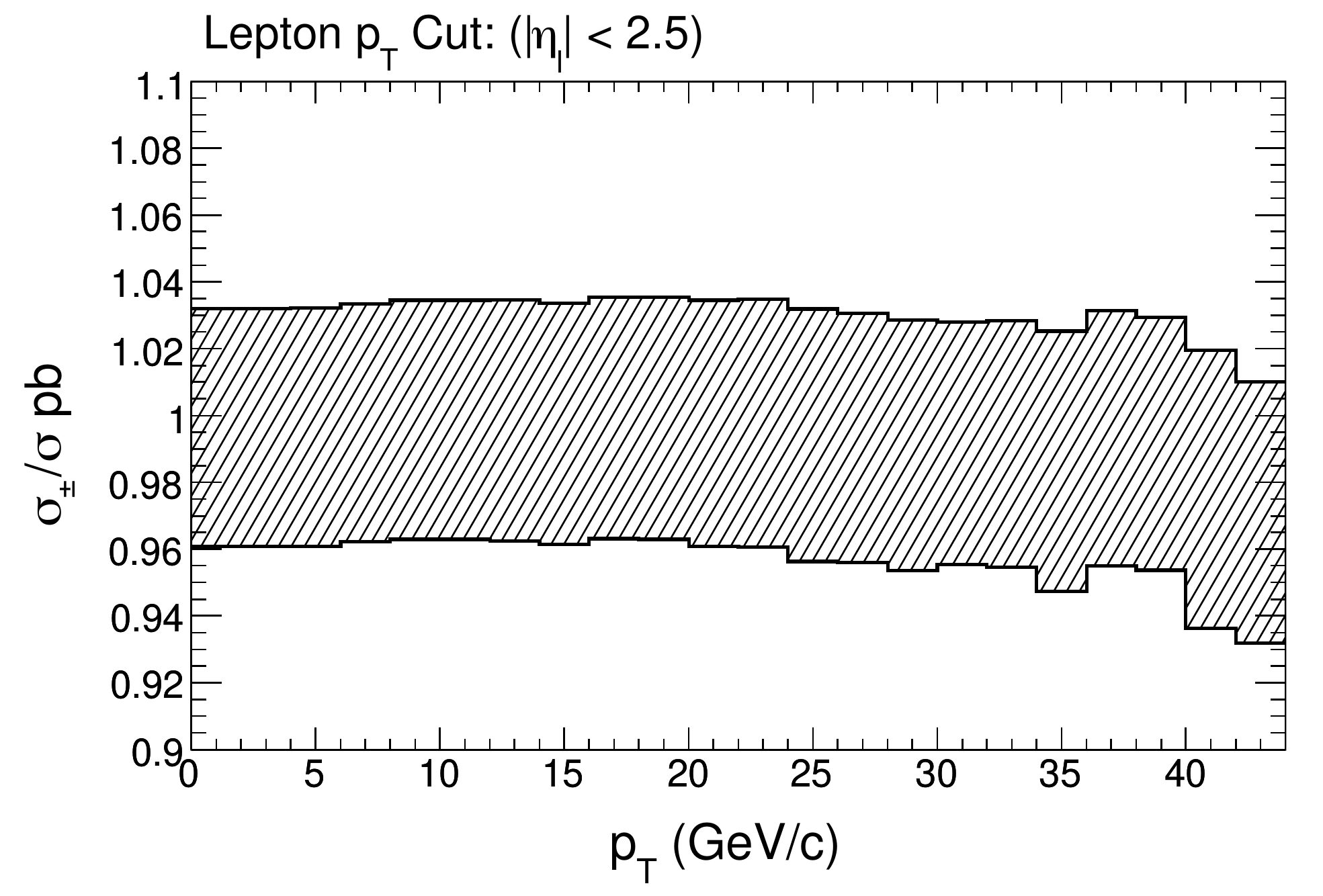} \\
\multicolumn{2}{c}{(a)} \\
\includegraphics[width=7cm]{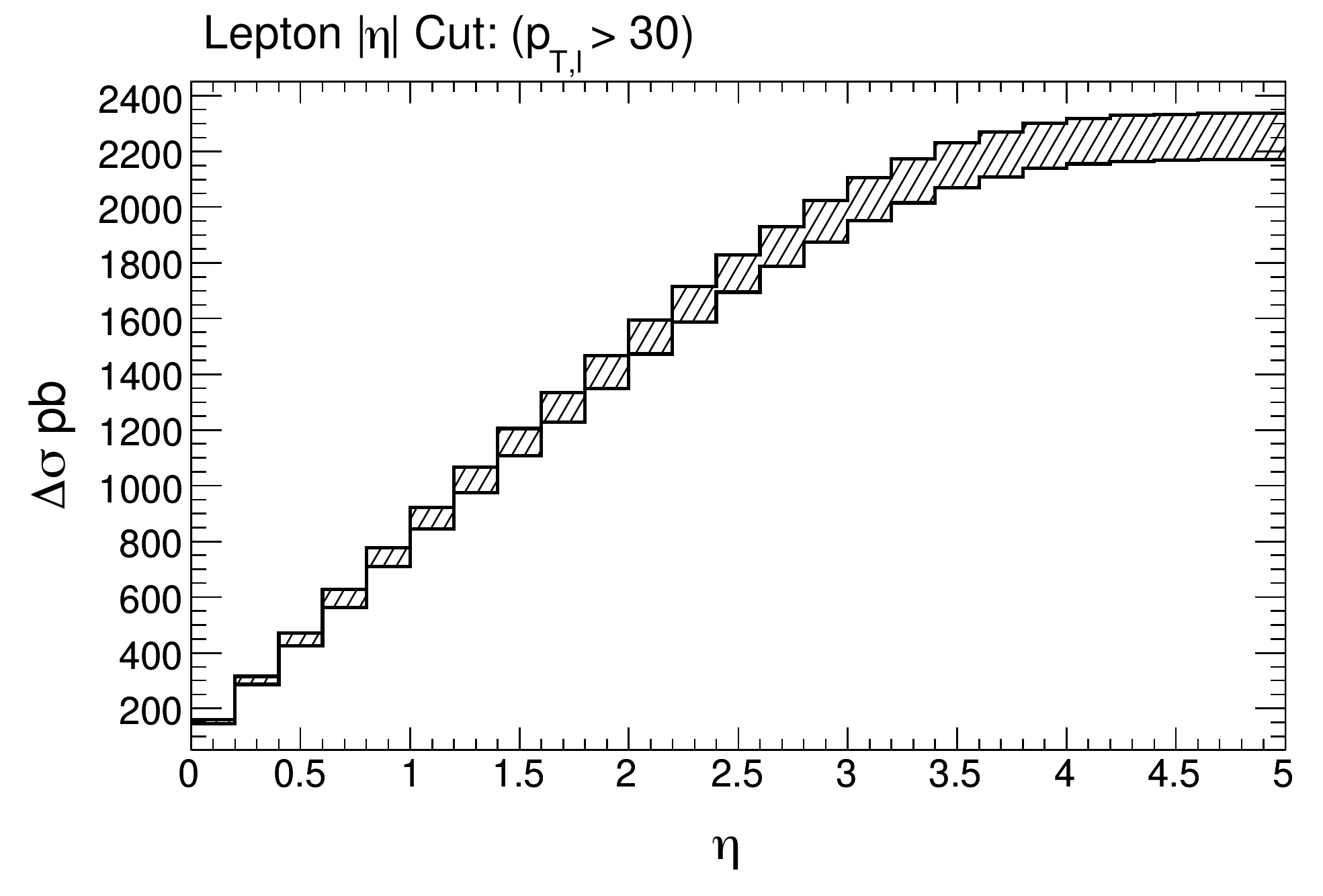}&
\includegraphics[width=7cm]{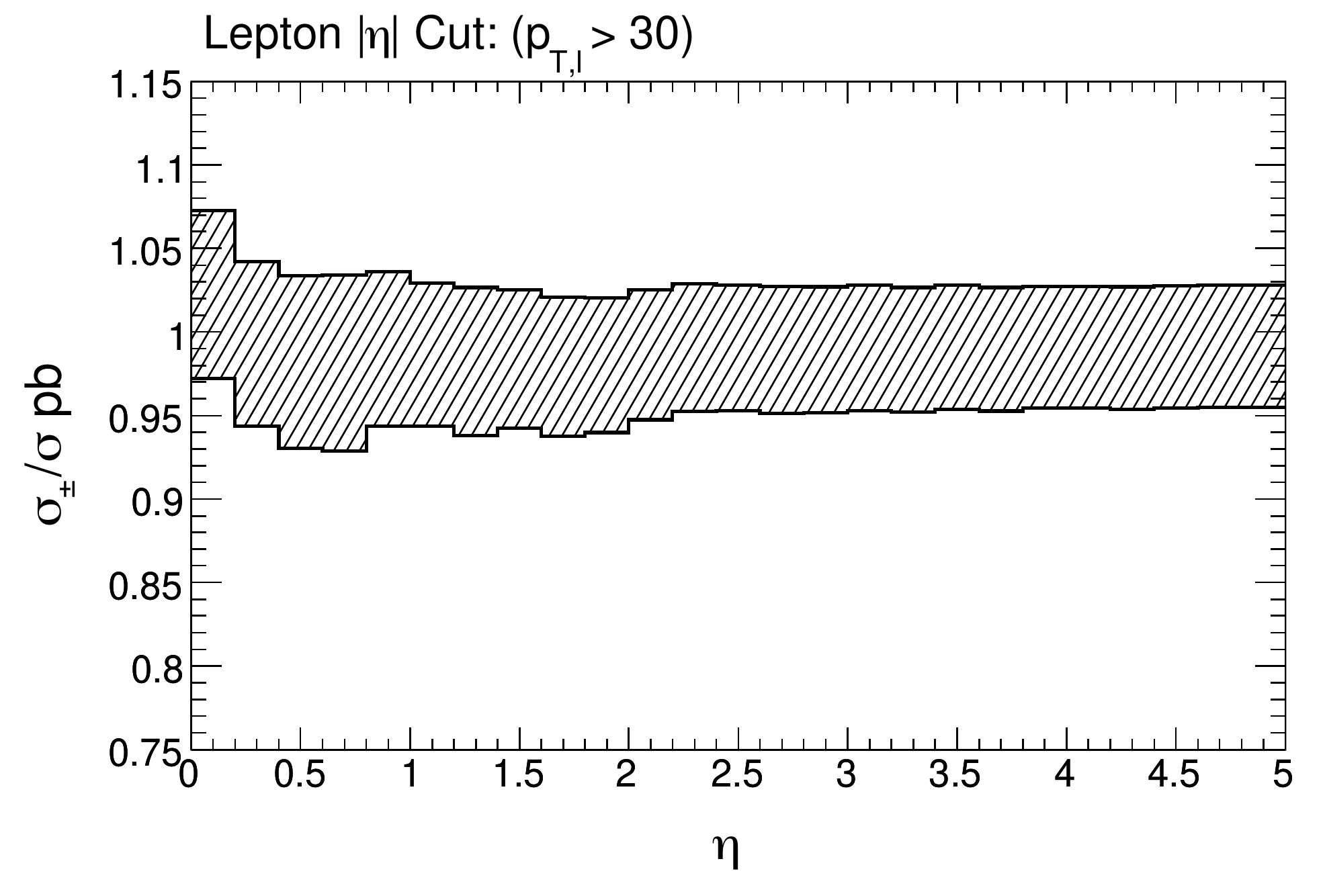}\\
\multicolumn{2}{c}{(b)} \\
\includegraphics[width=7cm]{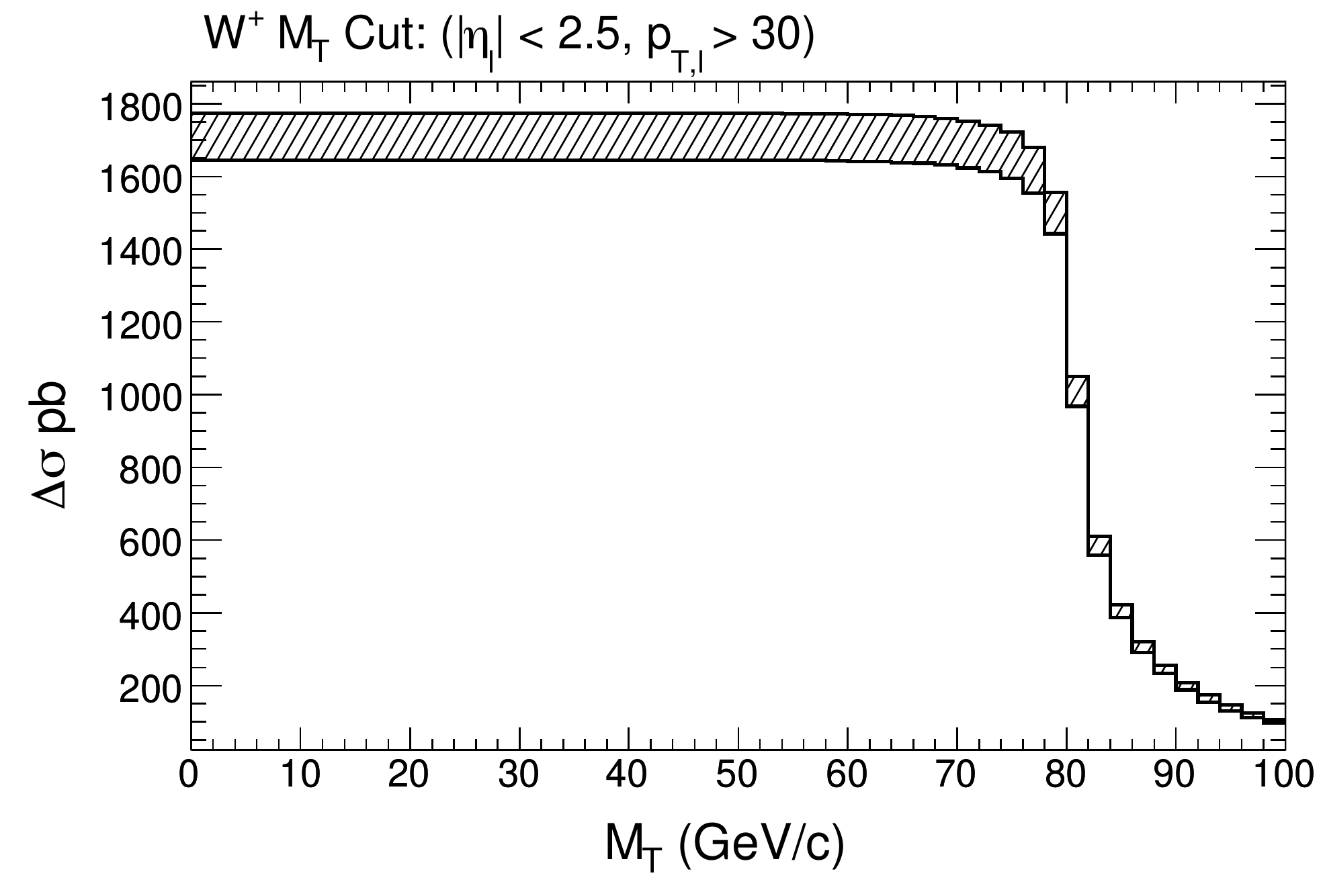}&
\includegraphics[width=7cm]{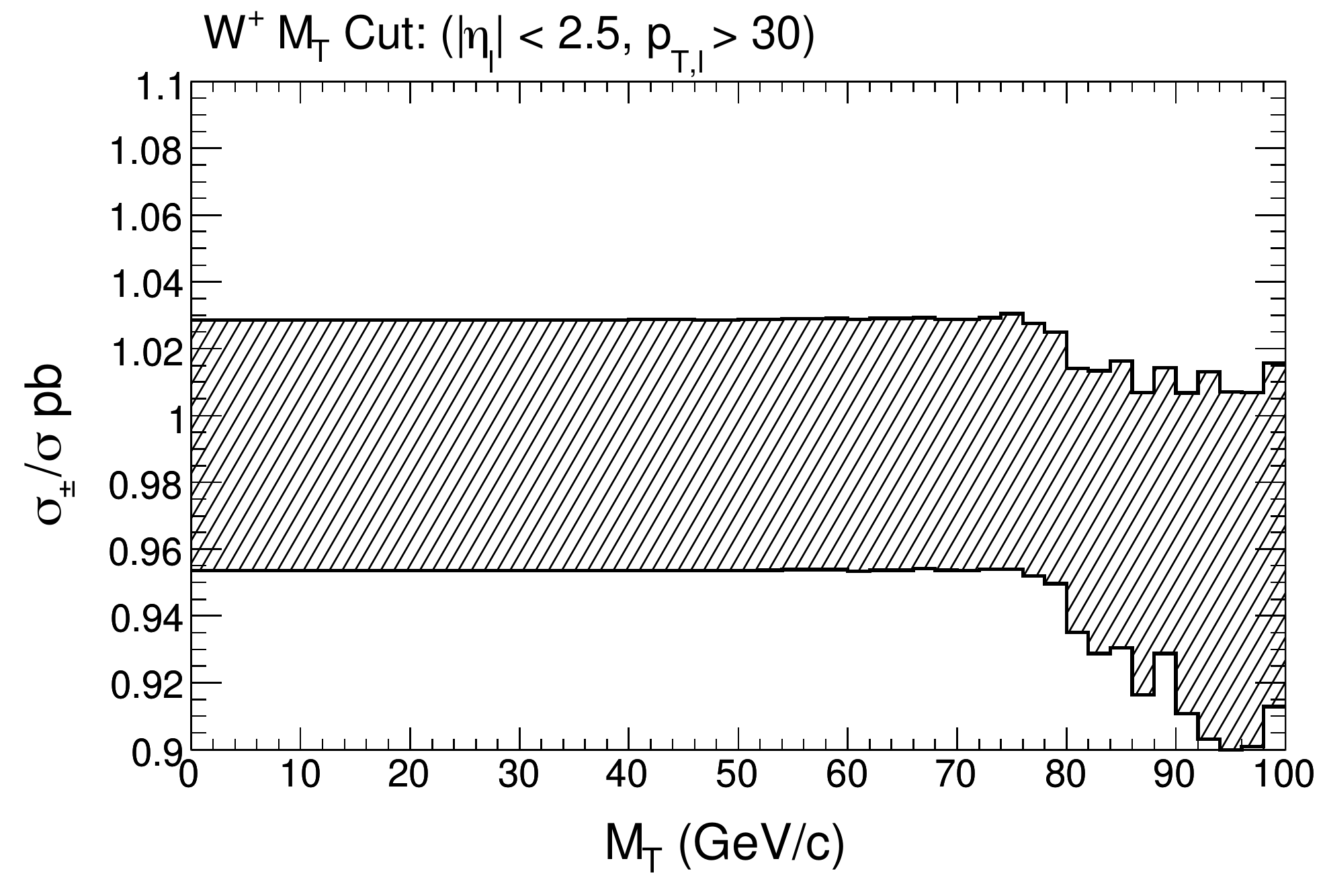}\\
\multicolumn{2}{c}{(c)} \\
\end{tabular}
\caption{ The $\Wml$ cross-section $\sigma$ at 7 TeV, as a
  function of the (a) $\pT$ cut, (b) $\eta$ cut, and (c) $M_{T}$ cut for acceptance regions
  as defined in Table~\ref{table:cuts}. We fix all cuts except the
  cut to be varied at their specified values. The figures on the right
  show the relative errors in the cross sections.}
\label{fig:pdf_xs_vs_cut_Wplus}
}

\FIGURE[ht]{
\begin{tabular}{cc}
\includegraphics[width=7cm]{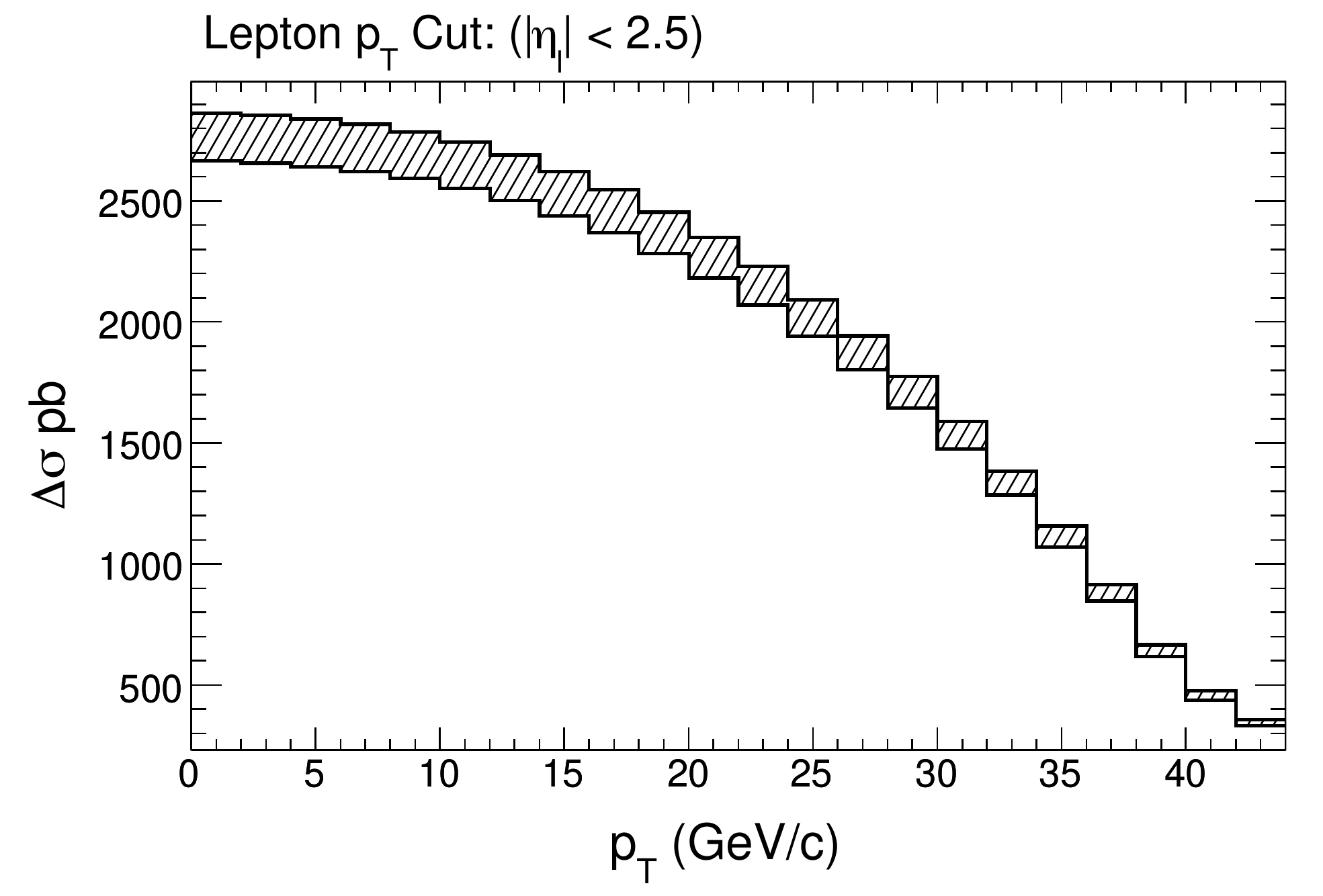} &
\includegraphics[width=7cm]{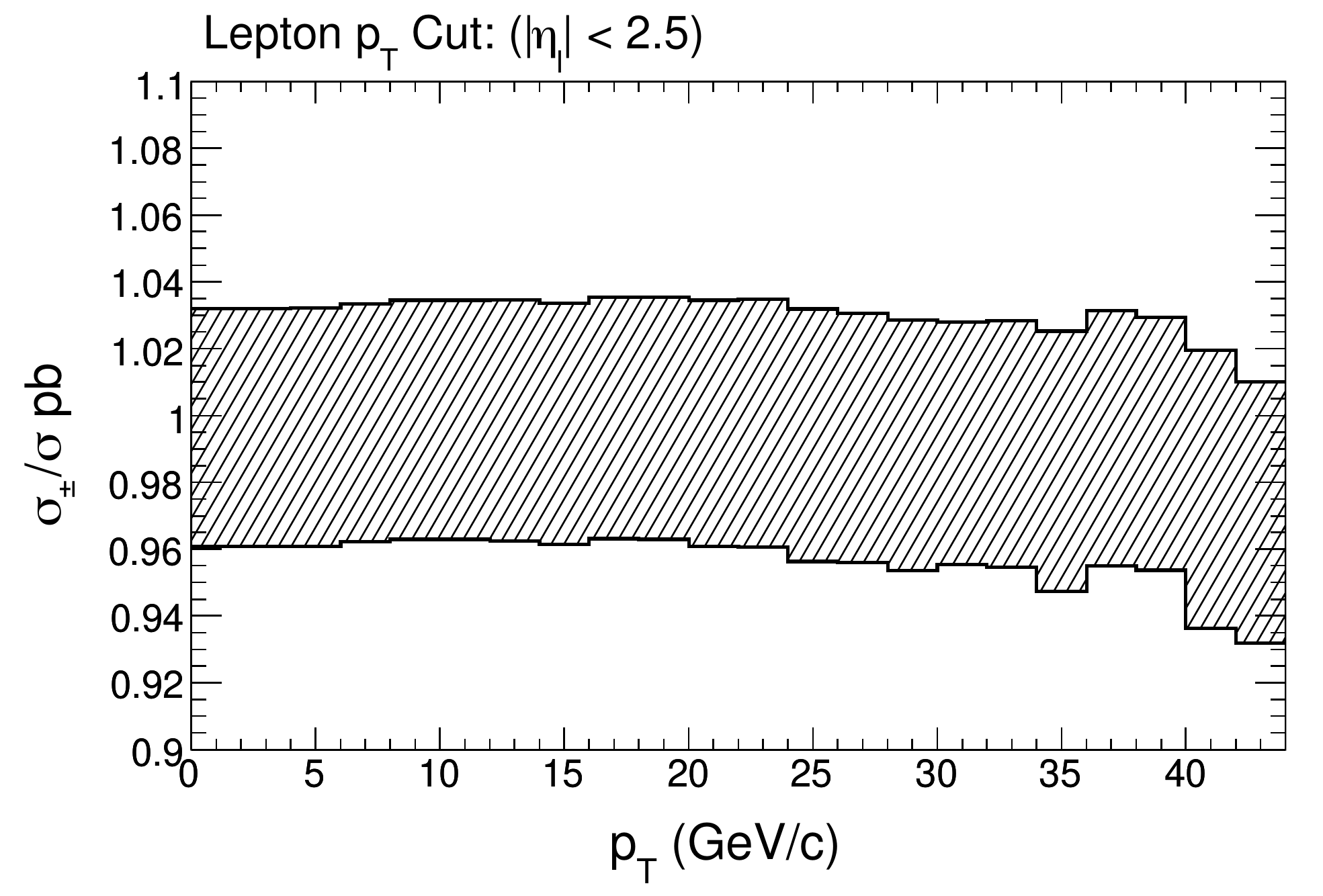} \\
\multicolumn{2}{c}{(a)} \\
\includegraphics[width=7cm]{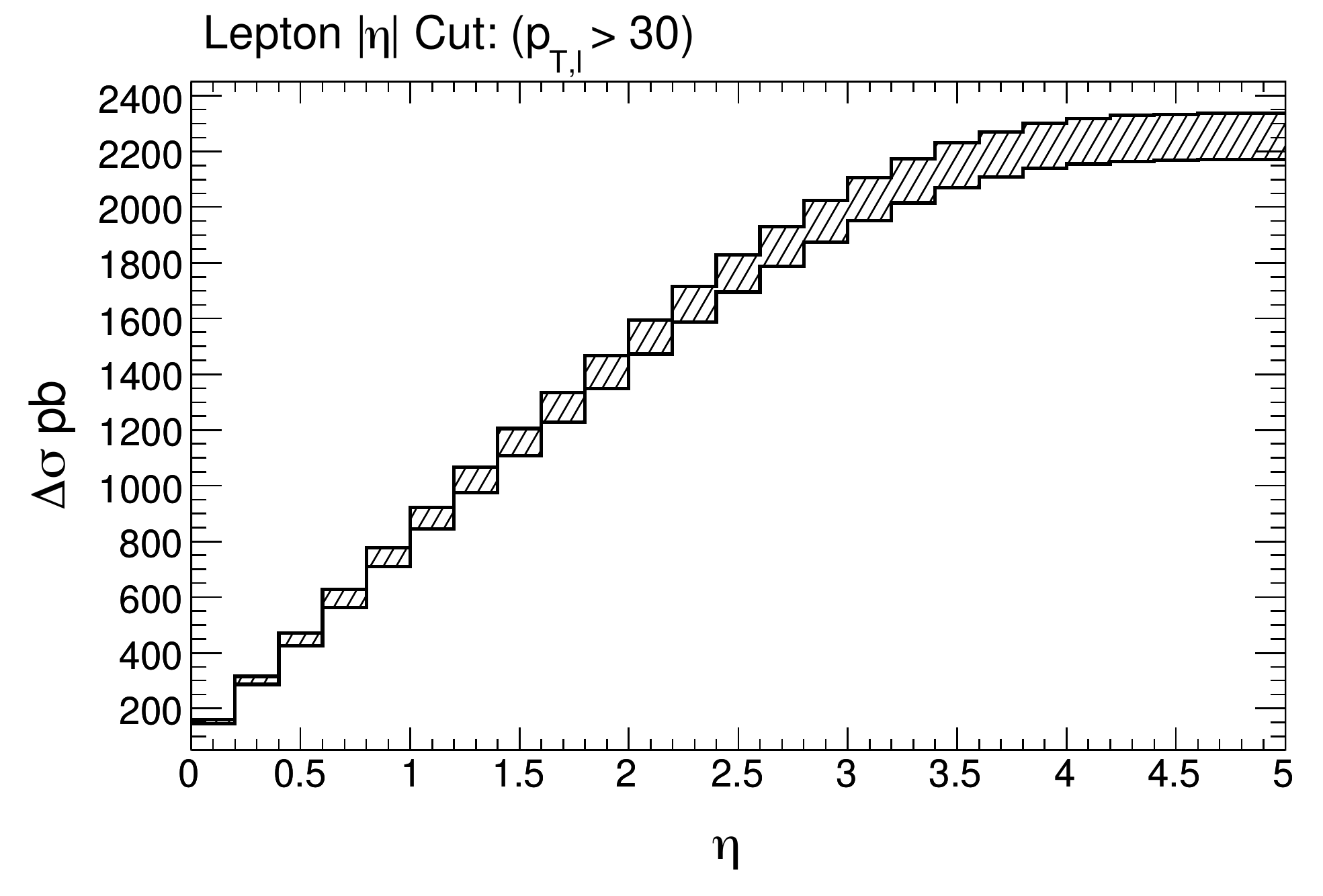}&
\includegraphics[width=7cm]{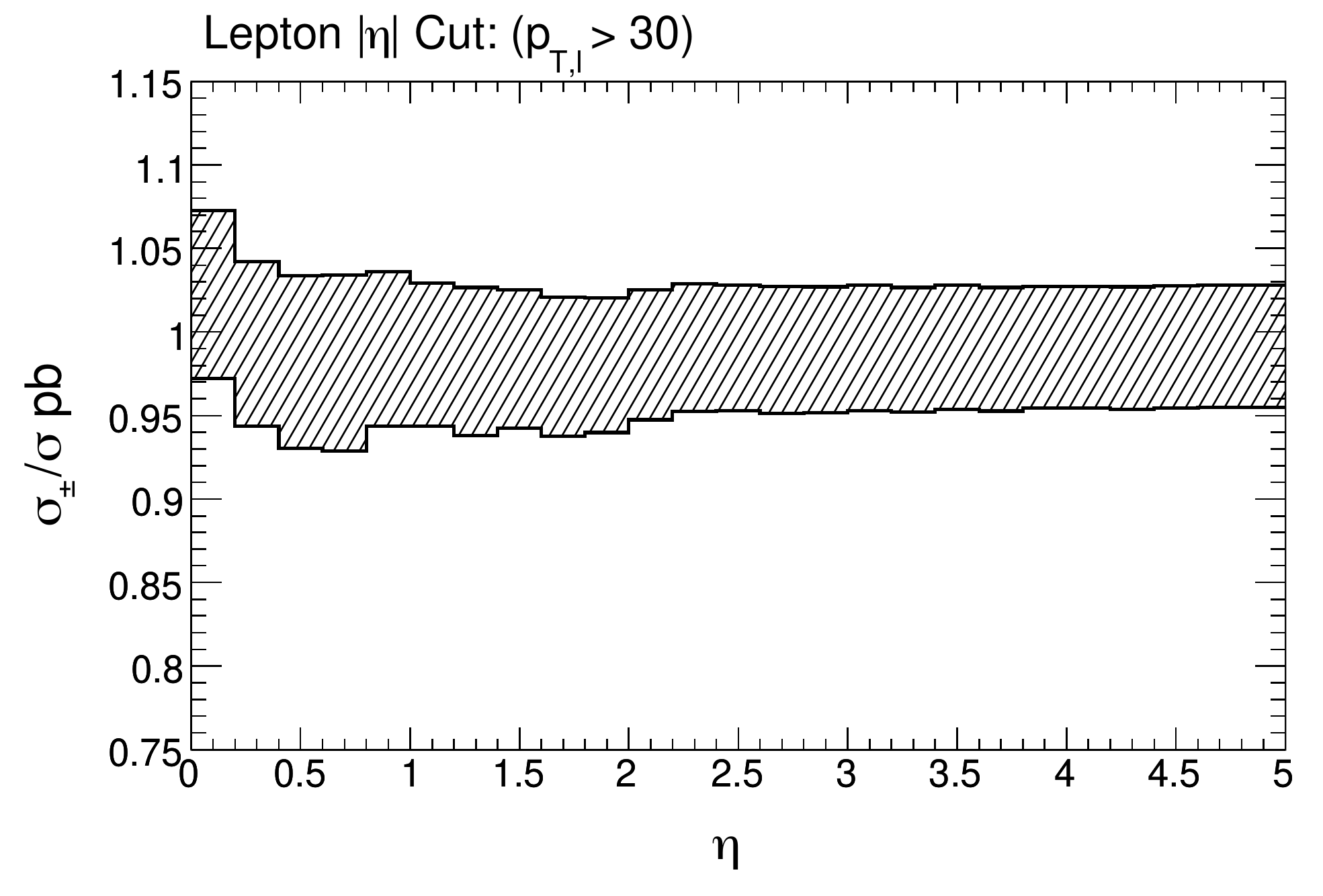}\\
\multicolumn{2}{c}{(b)} \\
\includegraphics[width=7cm]{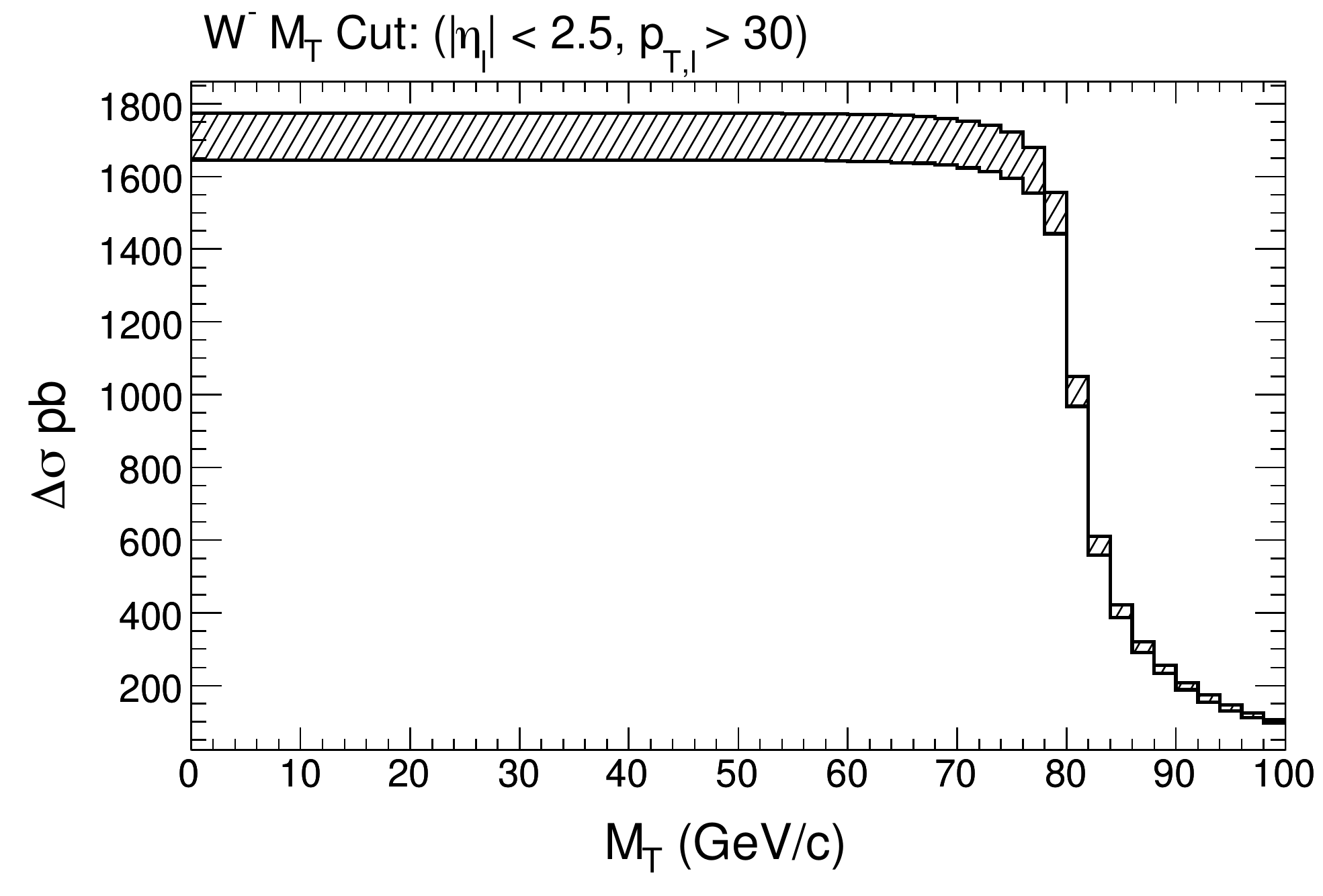}&
\includegraphics[width=7cm]{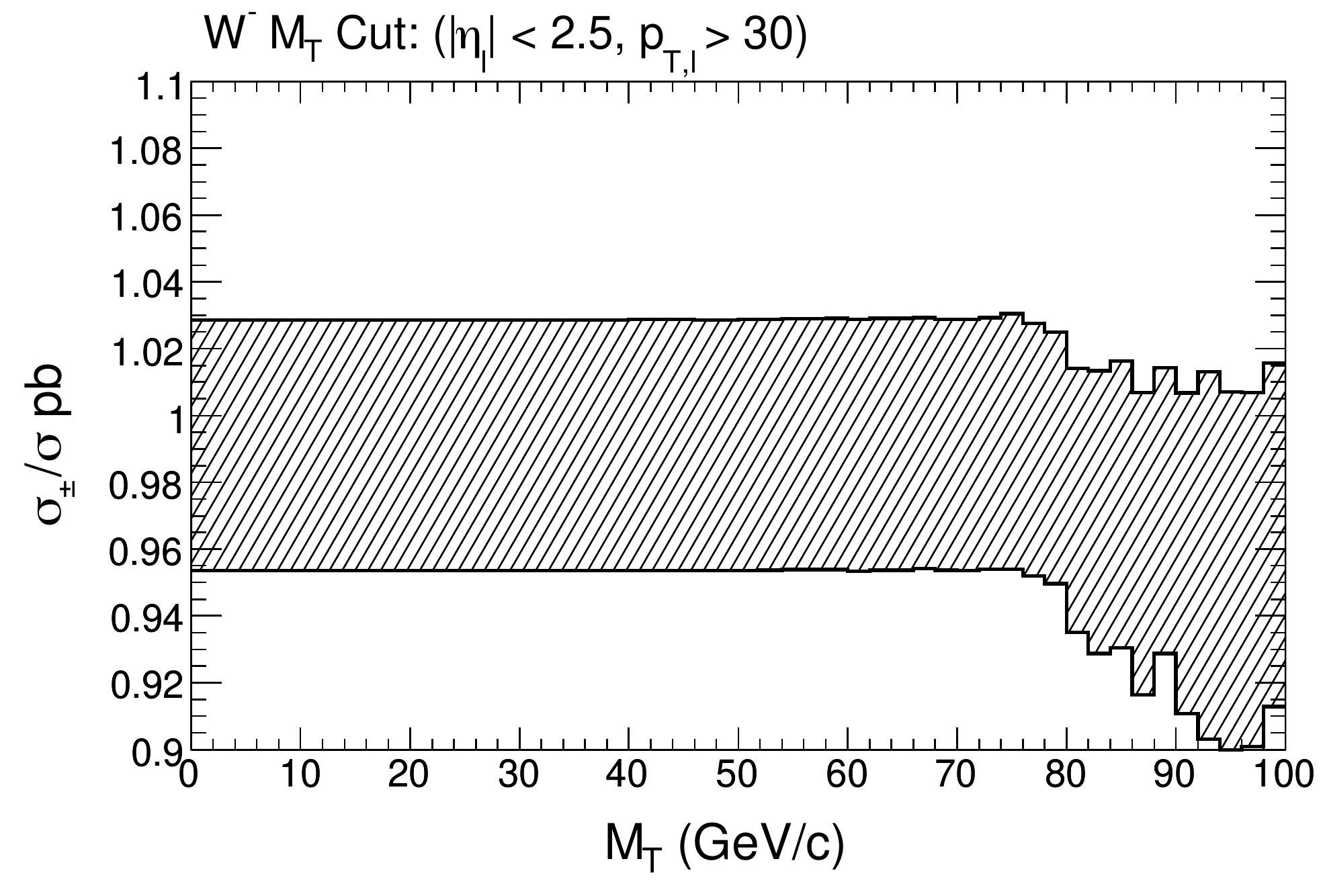}\\
\multicolumn{2}{c}{(c)} \\
\end{tabular}
\caption{ The $\Wml$ cross-section $\sigma$ at 7 TeV, as a
  function of the (a) $\pT$ cut, (b) $\eta$ cut, and (c) $M_{T}$ cut for acceptance regions
  as defined in Table~\ref{table:cuts}. We fix all cuts except the
  cut to be varied at their specified values. The figures on the right
  show the relative errors in the cross sections.}
\label{fig:pdf_xs_vs_cut_Wminus}
}

\FIGURE[t]{
\begin{tabular}{cc}
\includegraphics[width=7cm]{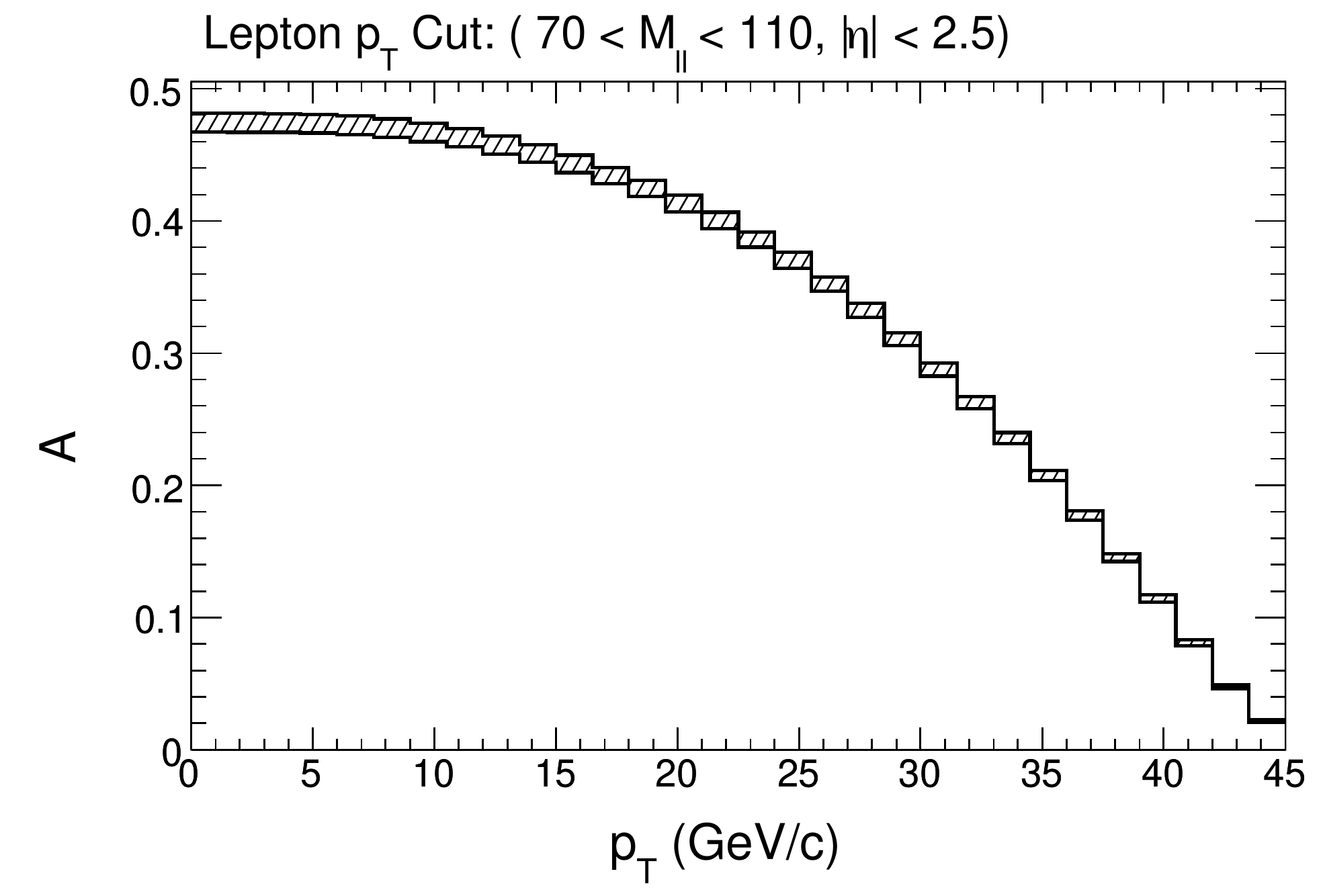} &
\includegraphics[width=7cm]{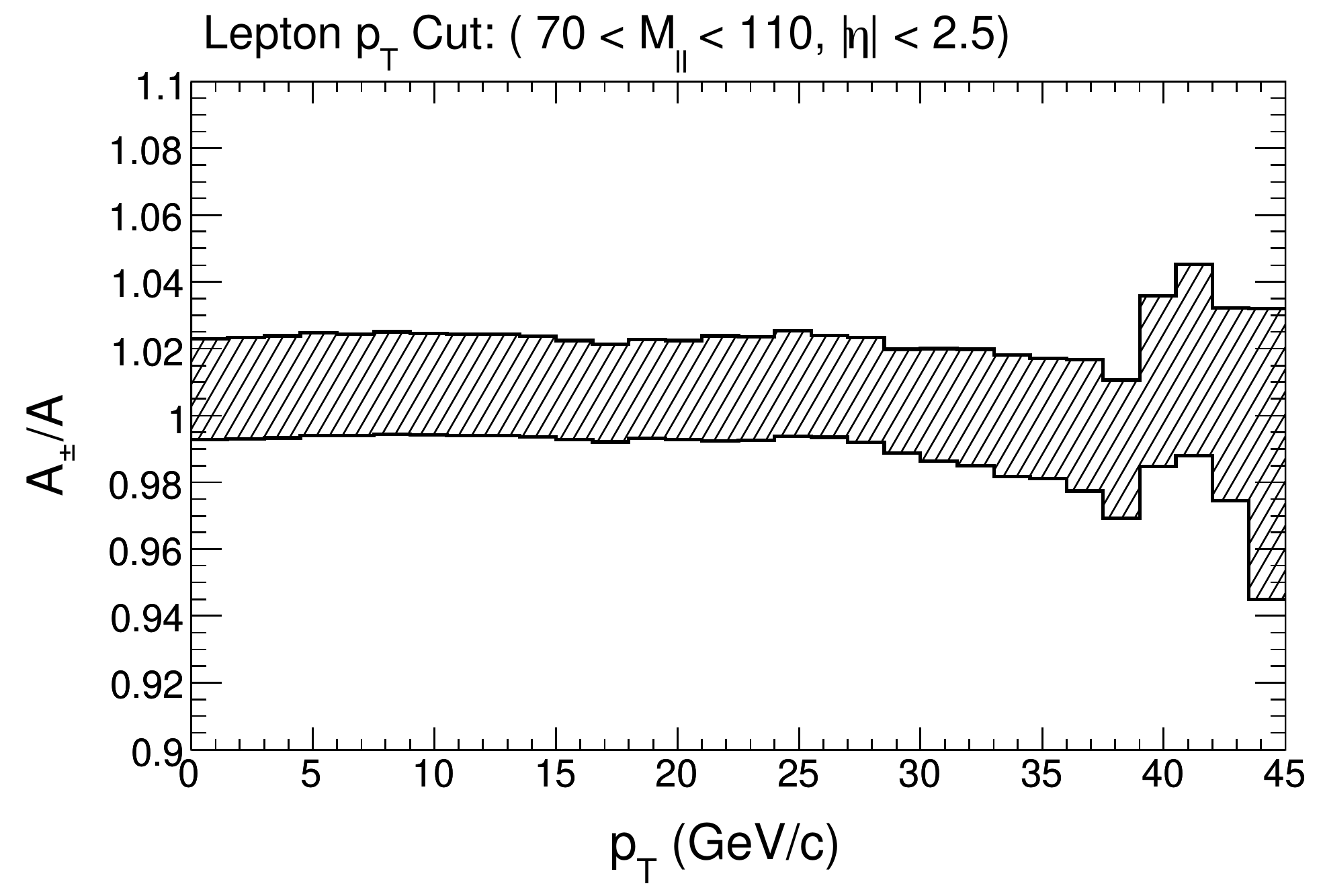} \\
\multicolumn{2}{c}{(a)} \\
\includegraphics[width=7cm]{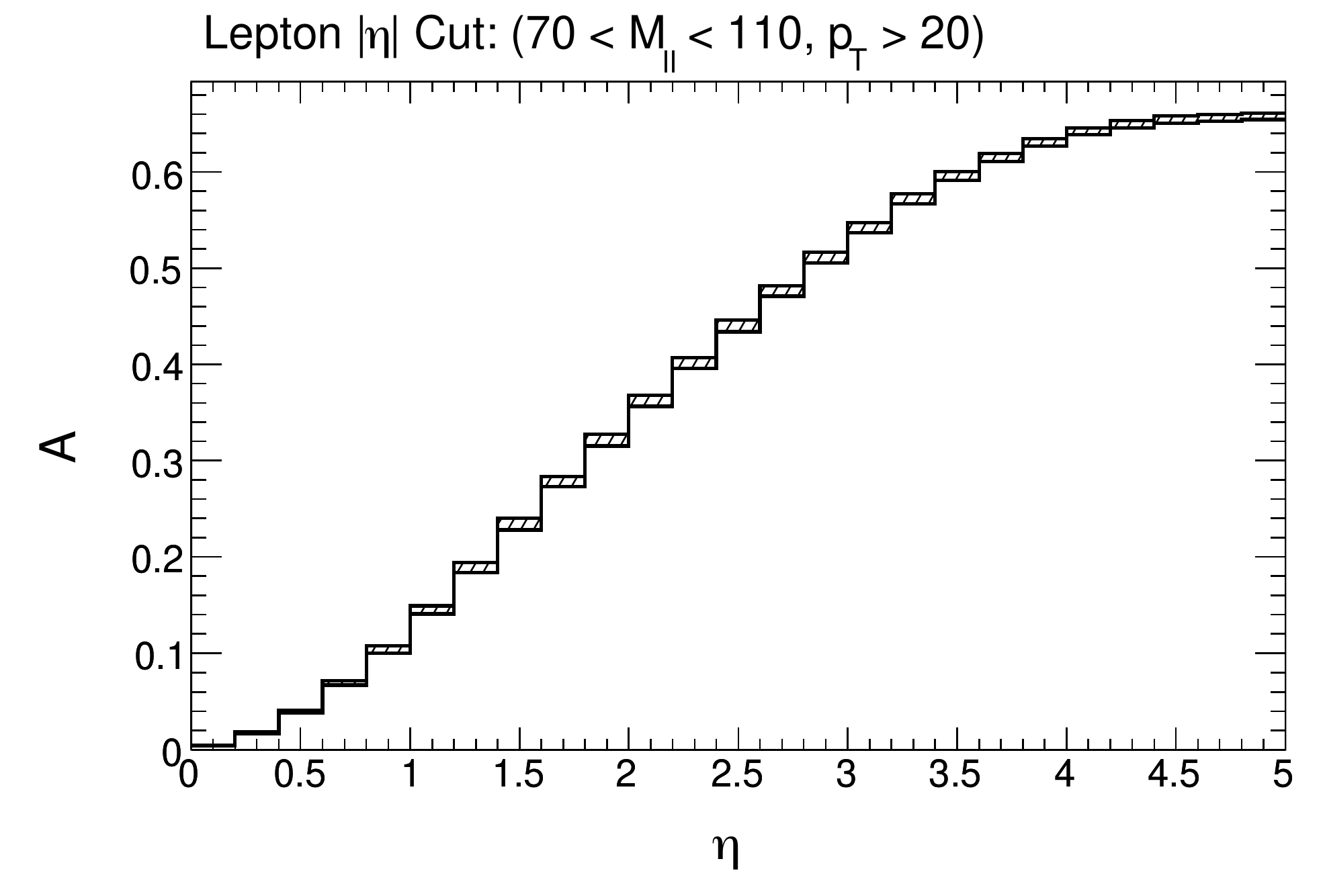}&
\includegraphics[width=7cm]{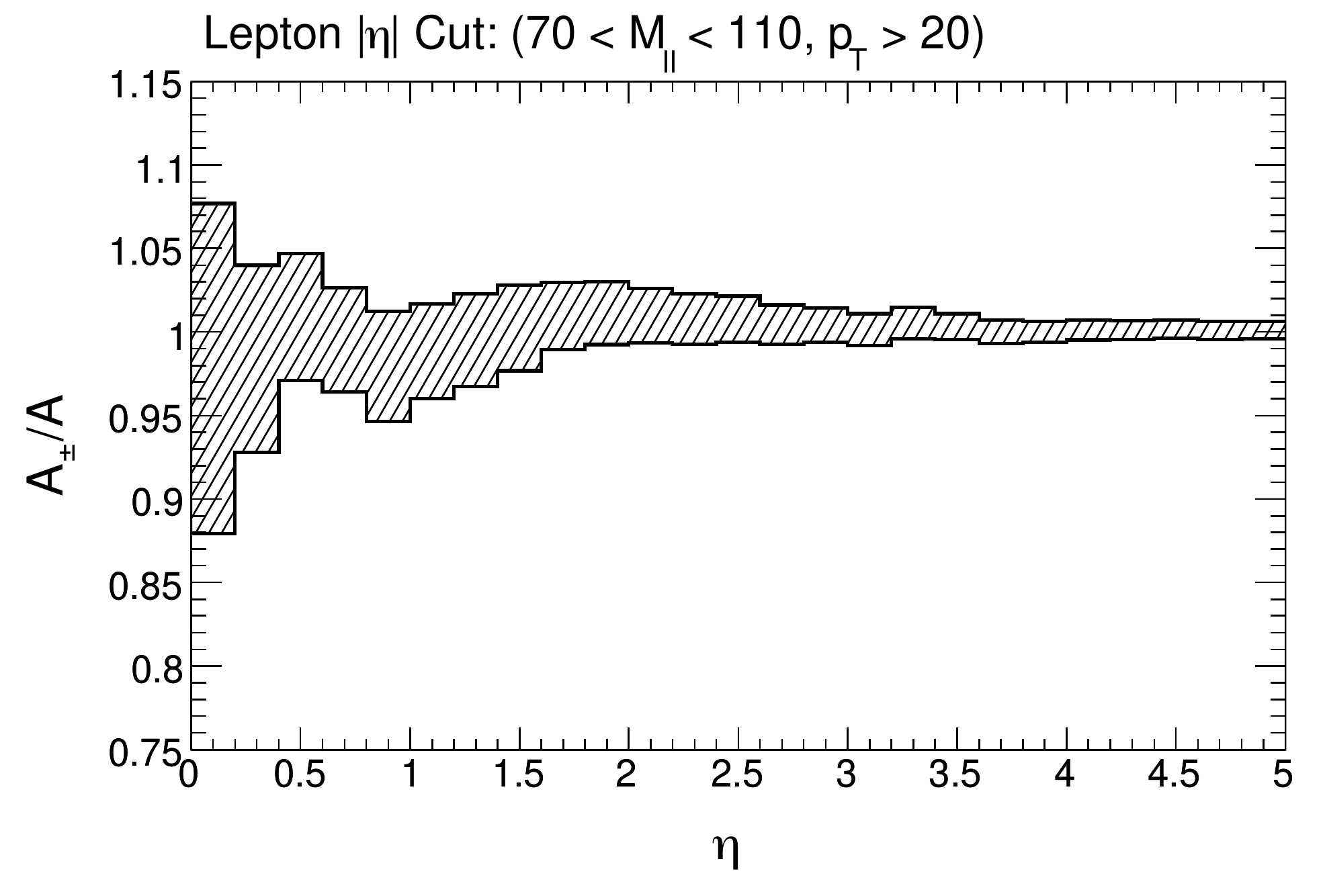}\\
\multicolumn{2}{c}{(b)} \\
\end{tabular}
\caption{ The $Z/\gamma^* \to \ell^{+}\ell^{-}$ acceptances $A$ at 7TeV,  as a
  function of the (a) $\pT$ cut and (b) $\eta$ cut for acceptance regions
  as defined in Table~\ref{table:cuts}. We fix all cuts except the
  cut to be varied at their specified values. The figures on the right
  show the relative errors in the acceptances.}
\label{fig:pdf_acc_vs_cut_Z}
}

\FIGURE[t]{
\begin{tabular}{cc}
\includegraphics[width=7cm]{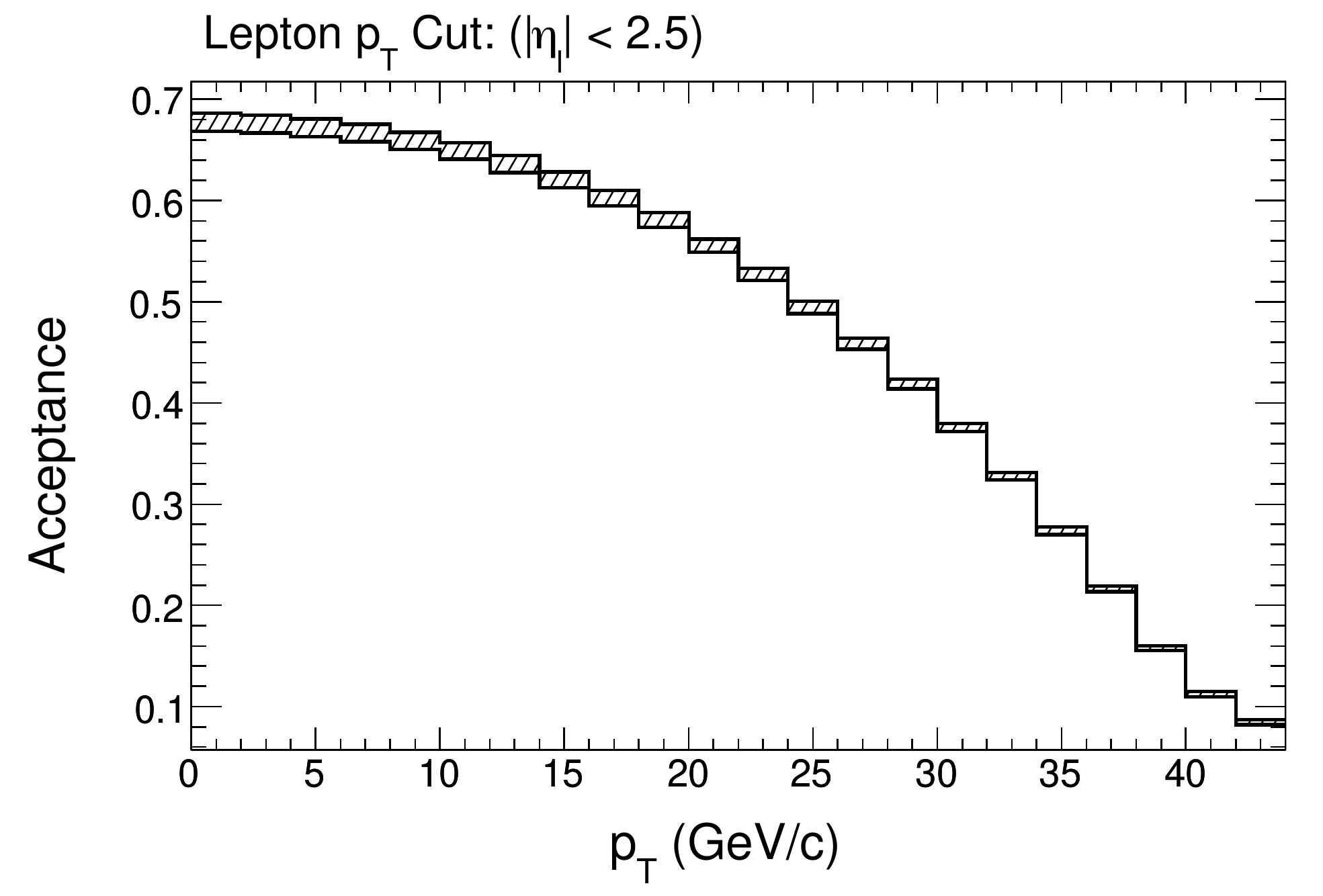} &
\includegraphics[width=7cm]{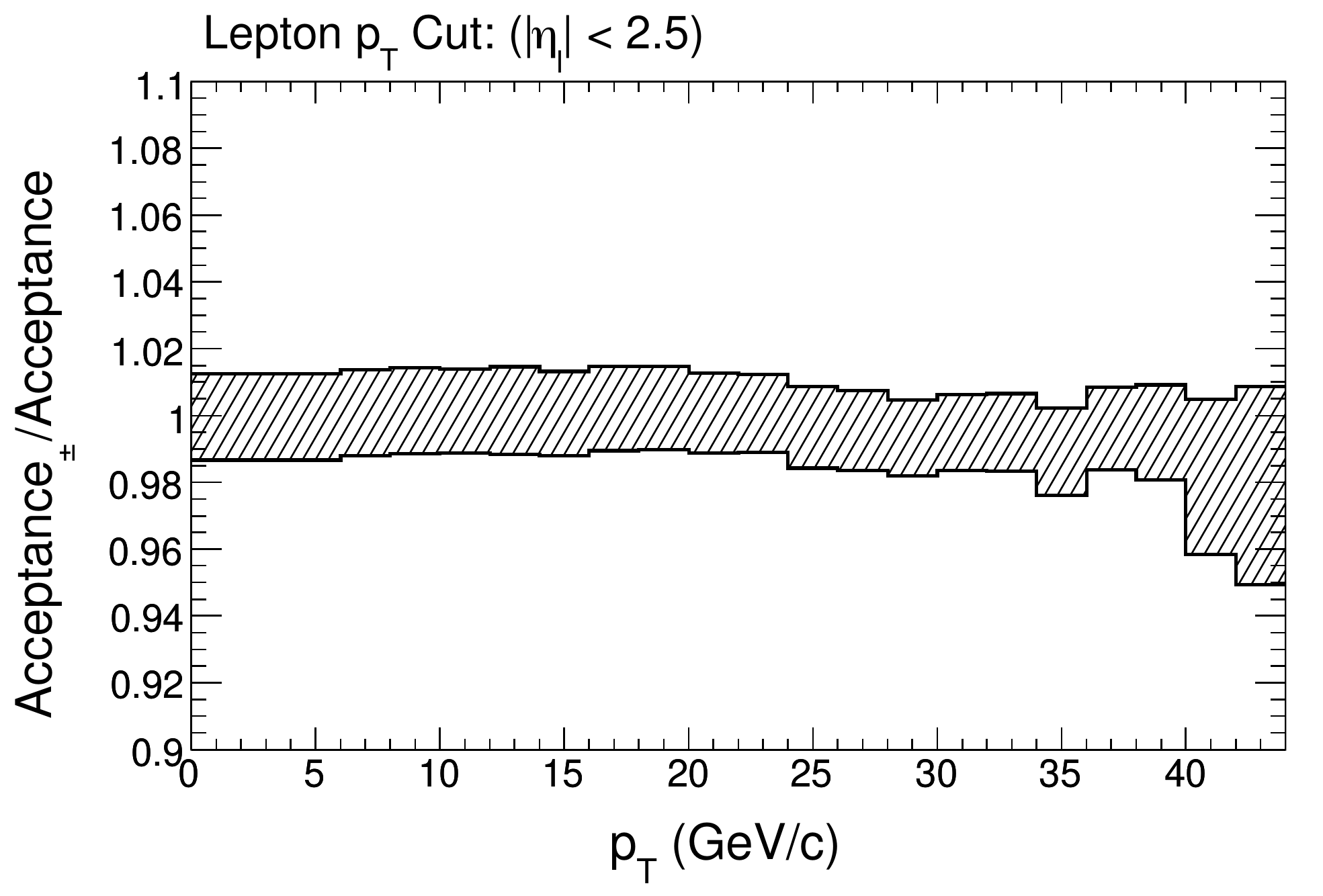} \\
\multicolumn{2}{c}{(a)} \\
\includegraphics[width=7cm]{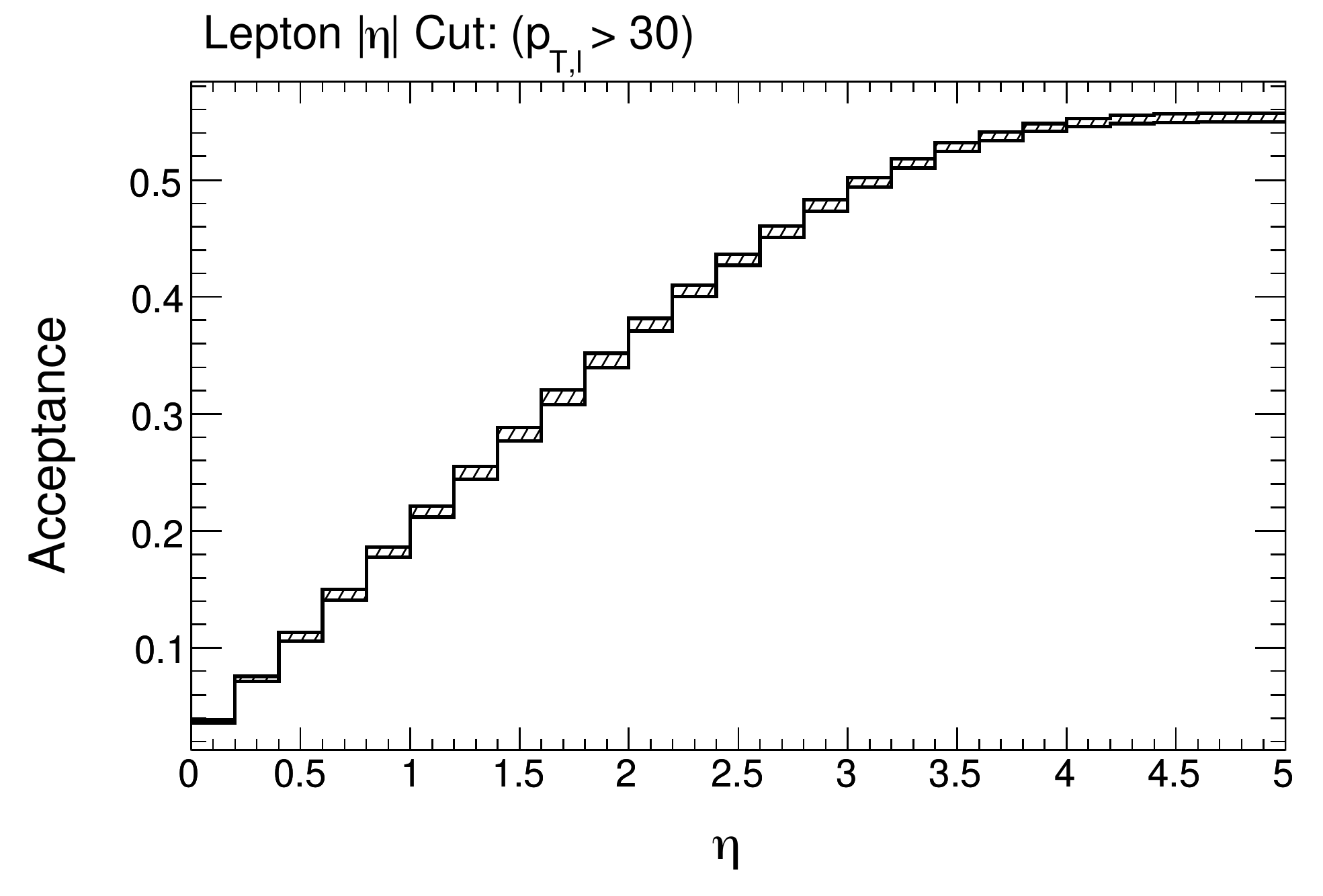}&
\includegraphics[width=7cm]{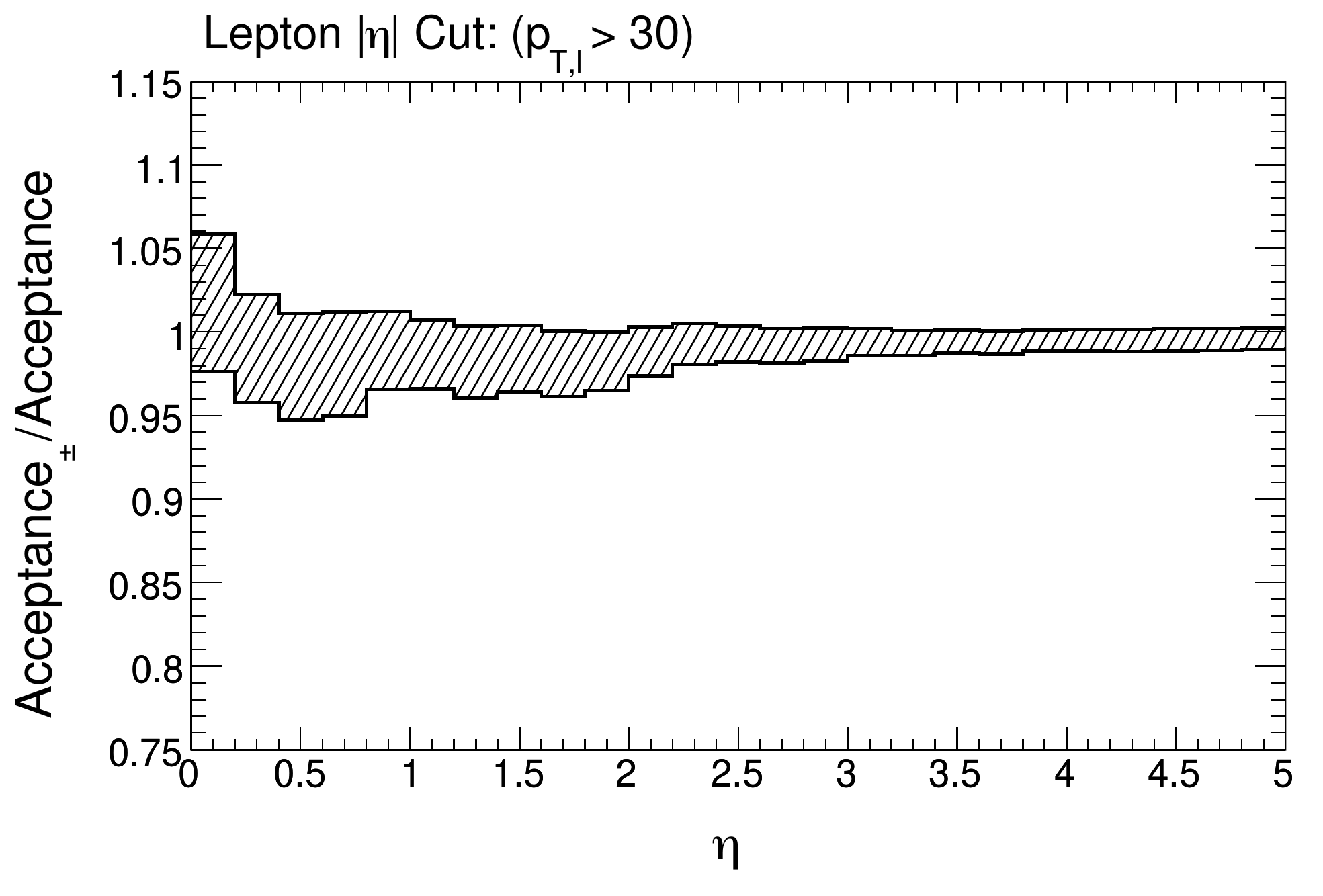}\\
\multicolumn{2}{c}{(b)} \\
\includegraphics[width=7cm]{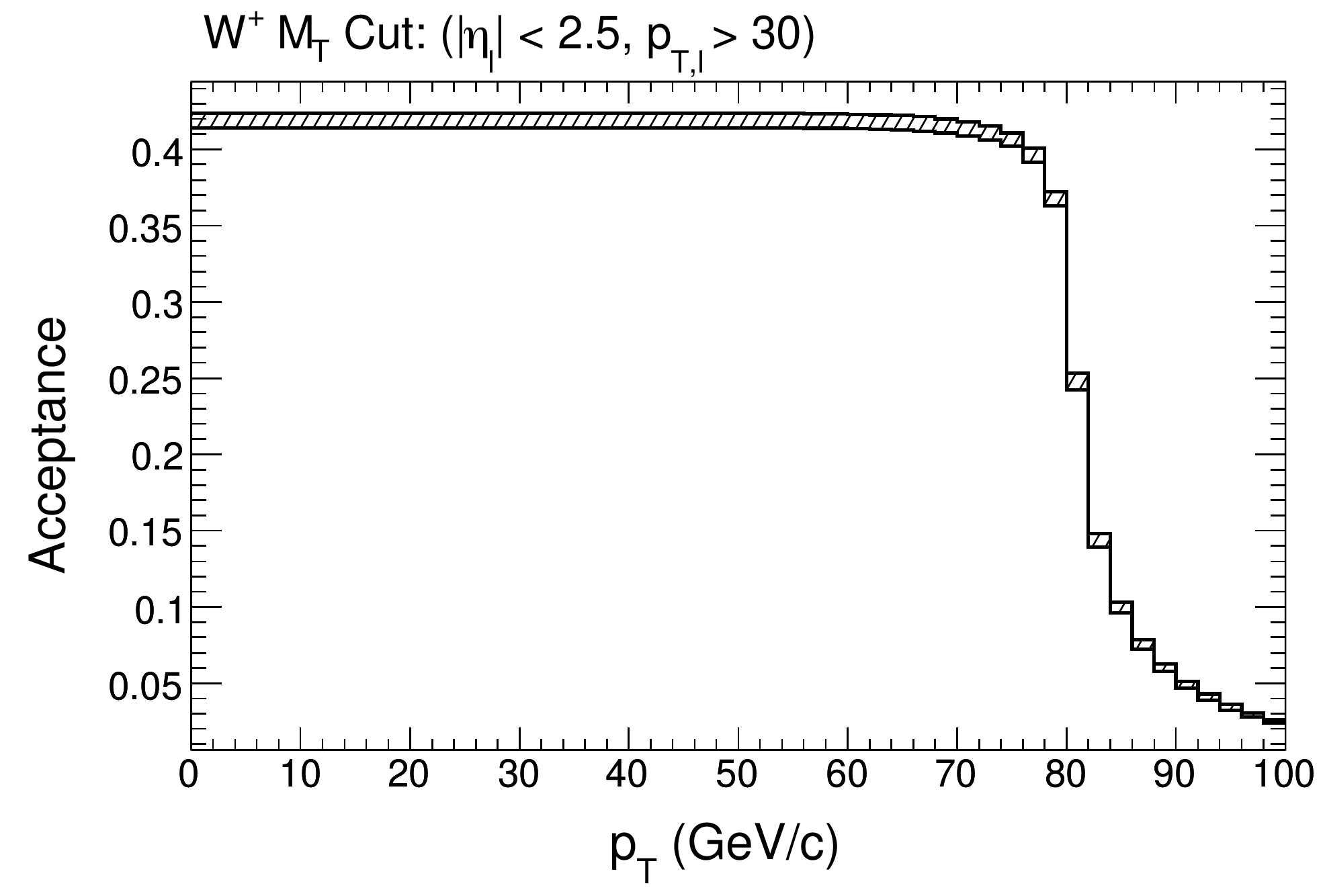}&
\includegraphics[width=7cm]{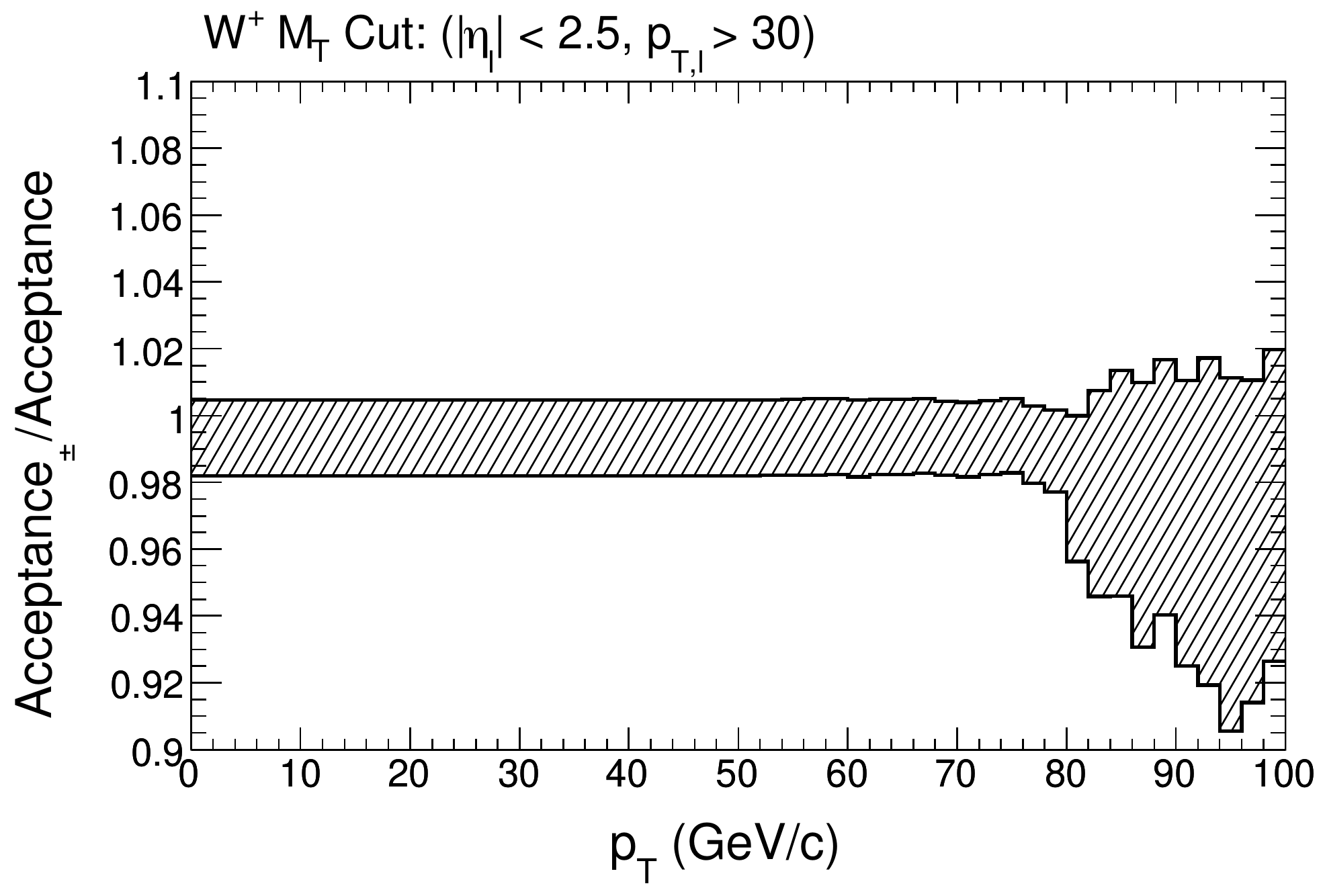}\\
\multicolumn{2}{c}{(c)} \\
\end{tabular}
\caption{ The $\Wpl$ acceptances $A$ at 7TeV,  as a
  function of the (a) $\pT$ cut, (b) $\eta$ cut, and (c) $M_{T}$ cut for acceptance regions
  as defined in Table~\ref{table:cuts}. We fix all cuts except the
  cut to be varied at their specified values. The figures on the right
  show the relative errors in the acceptances.}
\label{fig:pdf_acc_vs_cut_Wplus}
}

\FIGURE[t]{
\begin{tabular}{cc}
\includegraphics[width=7cm]{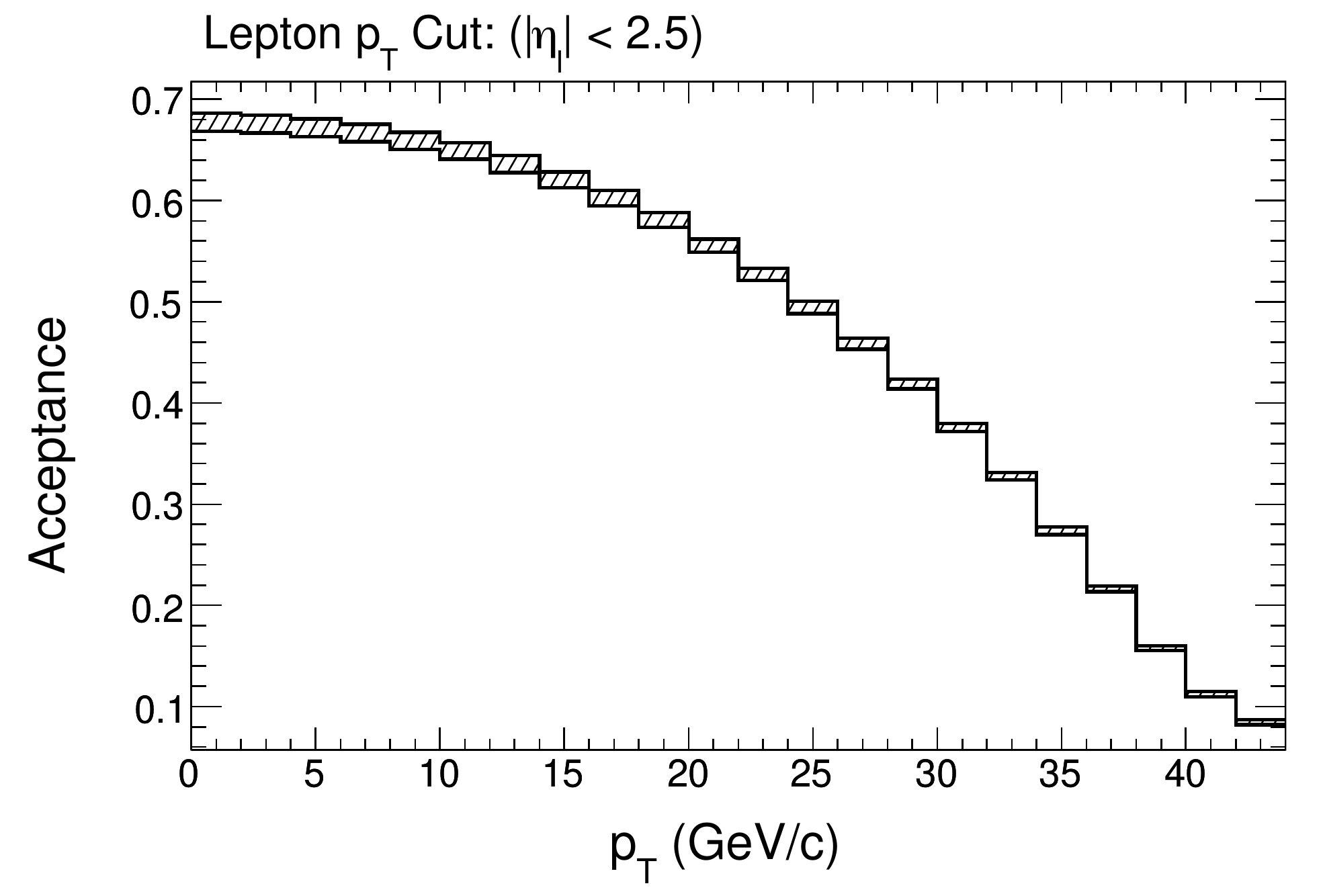} &
\includegraphics[width=7cm]{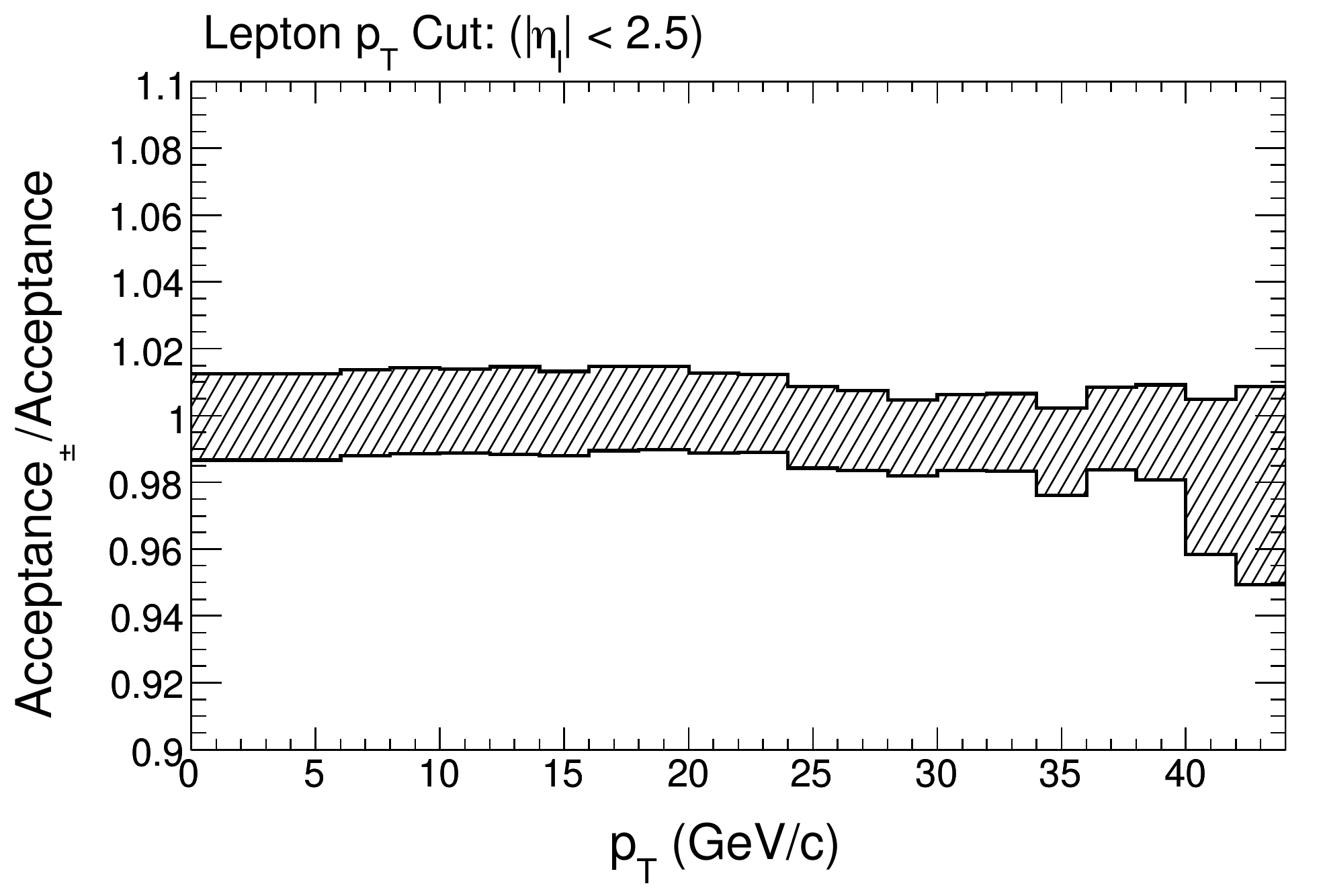} \\
\multicolumn{2}{c}{(a)} \\
\includegraphics[width=7cm]{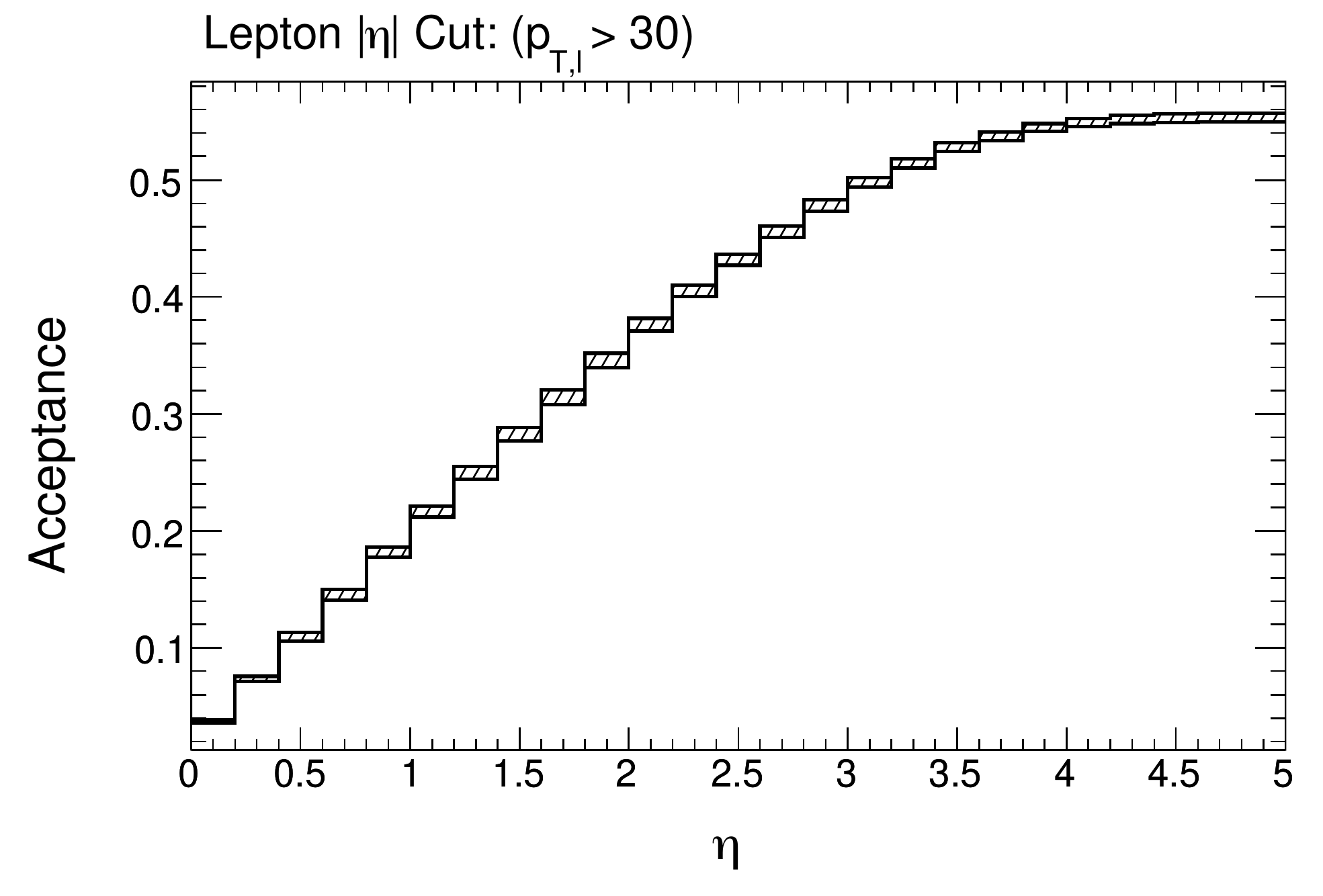}&
\includegraphics[width=7cm]{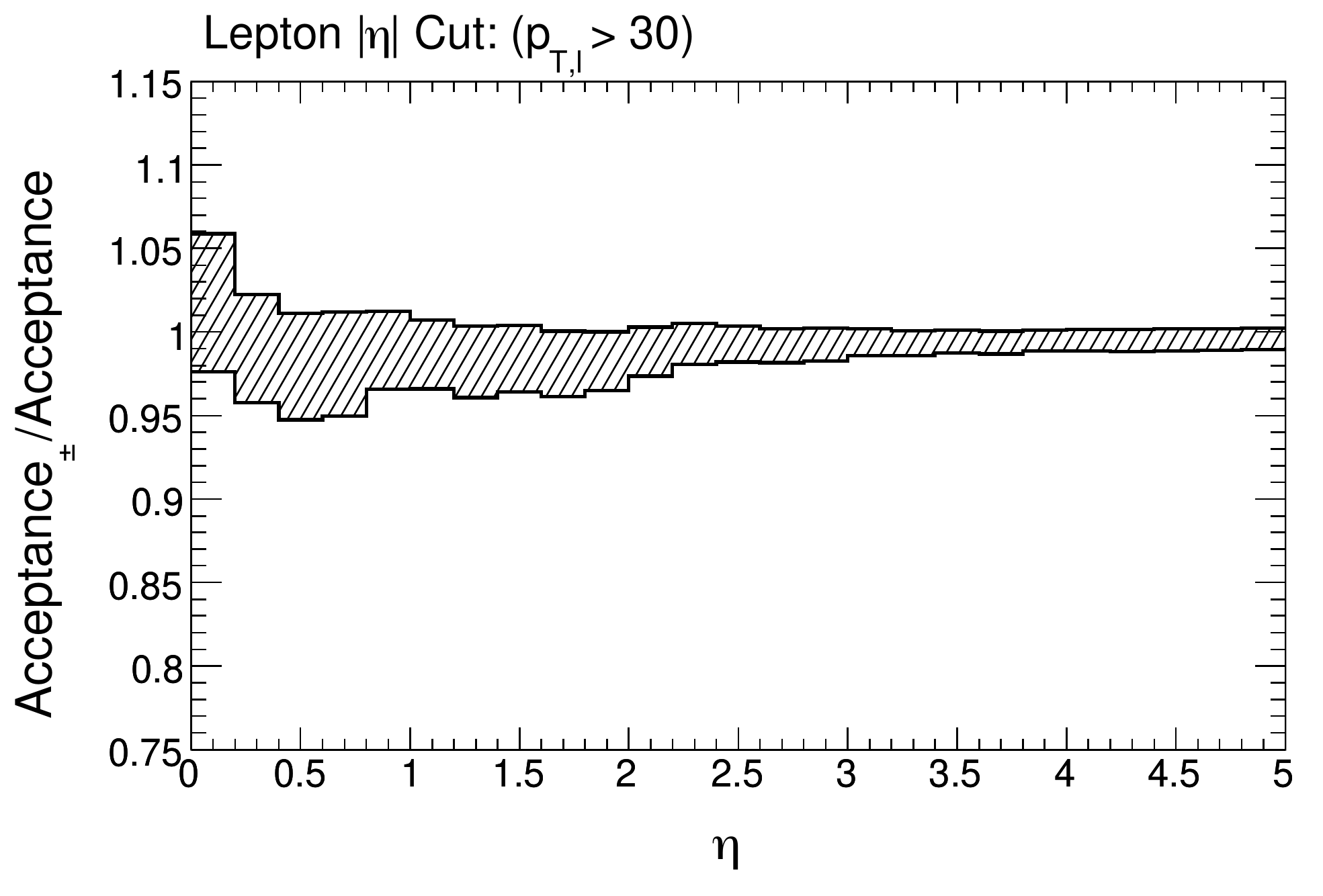}\\
\multicolumn{2}{c}{(b)} \\
\includegraphics[width=7cm]{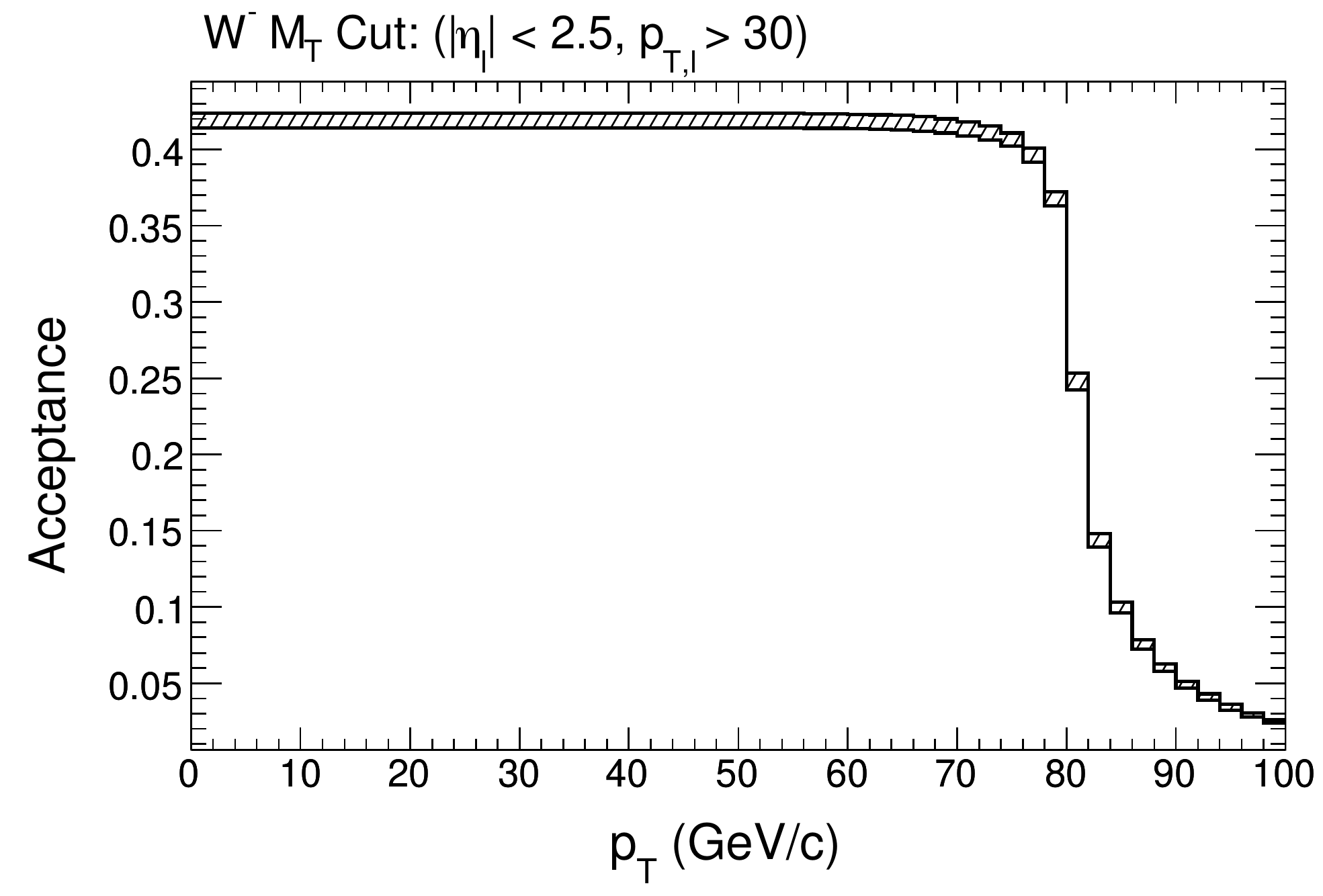}&
\includegraphics[width=7cm]{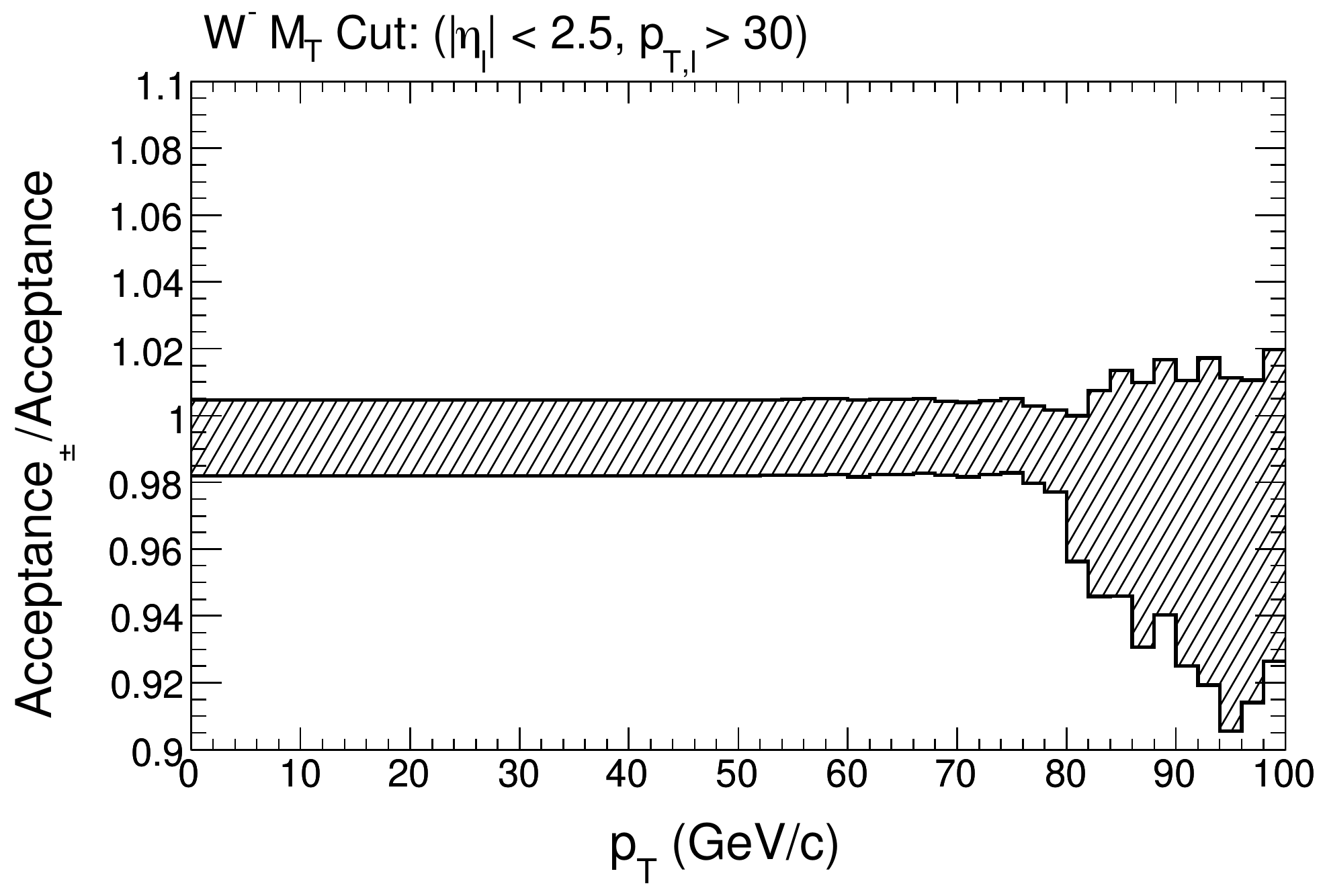}\\
\multicolumn{2}{c}{(c)} \\
\end{tabular}
\caption{ The $\Wml$ acceptances $A$ at 7TeV,  as a
  function of the (a) $\pT$ cut, (b) $\eta$ cut, and (c) $M_{T}$ cut for acceptance regions
  as defined in Table~\ref{table:cuts}. We fix all cuts except the
  cut to be varied at their specified values. The figures on the right
  show the relative errors in the acceptances.}
\label{fig:pdf_acc_vs_cut_Wminus}
}

\vfill
\end{document}